\documentclass[fleqn,usenatbib,useAMS]{mnras}

\usepackage{newtxtext,newtxmath}
\usepackage[T1]{fontenc}
\usepackage{ae,aecompl}
\usepackage{soul}
\usepackage{color}
\usepackage{soul}
\usepackage{graphicx}	
\usepackage{amsmath}	
\usepackage{booktabs}
\newcommand{\ra}[1]{\renewcommand{\arraystretch}{#1}}

\usepackage{newtxtext,newtxmath}
\title[The peculiar system NGC~1487]{Physical and kinematic conditions of the local merging galaxy NGC~1487}

\author[M. L. Buzzo et al.]{
M. L. Buzzo$^{1,2}$\thanks{E-mail: maria.buzzo@usp.br},
B. Ziegler$^{2}$,
P. Amram$^{3}$,
M. Verdugo$^{2}$,
C. E. Barbosa$^{1}$, 
B. Ciocan$^{2}$,
\newauthor
P. Papaderos$^{4,5,2}$, 
S. {Torres-Flores}$^{6}$, 
C. {Mendes de Oliveira}$^{1}$
\\\\
$^{1}$ Universidade de S\~ao Paulo, IAG, Rua do Mat\~ao 1226, Cidade Universit\'aria, S\~ao Paulo 05508-900,Brazil\\
$^{2}$ University of Vienna, Department of Astronomy,
              T\"urkenschanzstrasse 17, A-1180 Vienna\\
$^{3}$ Aix Marseille Univ, CNRS, CNES, LAM, Laboratoire d’Astrophysique de Marseille, Marseille, France\\
$^{4}$ Instituto de Astrof\'isica e Ci\^encias do Espa\c{c}o, Universidade de Lisboa, OAL, Tapada da Ajuda, PT1349-018 Lisboa, Portugal \\
$^{5}$ Departamento de Física, Faculdade de Ciências da Universidade de Lisboa, Edifício C8, Campo Grande, PT1749-016 Lisboa, Portugal \\
$^{6}$ Departamento de Astronomía, Universidad de La Serena, Av. Juan Cisternas 1200 Norte, La Serena, Chile\\
}

\date{Accepted 2021 February 08. Received 2021 February 05; in original form 2020 October 23}

\pubyear{2021}

\begin{document}
\label{firstpage}
\pagerange{\pageref{firstpage}--\pageref{lastpage}}
\maketitle

\begin{abstract}
We present optical VLT/MUSE integral field spectroscopy data of the merging galaxy NGC~1487. We use fitting techniques to study the ionized gas emission of this merger and its main morphological and kinematical properties. 
We measured flat and sometimes inverted oxygen abundance gradients in the subsystems composing NGC~1487, explained by metal mixing processes common in merging galaxies.  
We also measured widespread star-forming bursts, indicating that photoionisation by stars is the primary ionization source of the galaxy.
The kinematic map revealed a rotating pattern in the gas in the northern tail of the system, suggesting that the galaxy may be in the process of rebuilding a disc. 
The gas located in the central region has larger velocity dispersion ($\sigma\approx 50$ km s$^{-1}$) than the remaining regions, indicating kinematic heating, possibly owing to the ongoing interaction. Similar trends were, however, not observed in the stellar velocity-dispersion map, indicating that the galaxy has not yet achieved equilibrium, and the nebular and stellar components are still kinematically decoupled. Based on all our measurements and findings, and specially on the mass estimates, metallicity gradients and velocity fields of the system, we propose that NGC 1487 is the result of an ongoing merger event involving smallish dwarf galaxies within a group, in a pre-merger phase, resulting in a relic with mass and physical parameters similar to a dwarf galaxy. Thus, we may be witnessing the formation of a dwarf galaxy by merging of smaller clumps at z=0.
\end{abstract}




\section{Introduction}
\label{sec:intro}
During the last decades several studies have shown that interacting galaxies display lower central abundances than isolated galaxies at a given mass \citep{K06}. This ``dilution'' in the central abundances may be produced by gas inflows from the less metal enriched areas in the outskirts of the galaxy to its centre. In the same context, observations have shown that interacting galaxies display flatter metallicity distributions than non-interacting objects \citep{torres-flores14,torres-flores15,Chien07,2015olave}, which is consistent with the above scenario, if we consider that flat gradients may be associated with gas flows of fresh gas from the outskirts inwards. 
However, only a couple of studies have confirmed observationally the connection between gas flows and flat metallicity gradients (e.g. \citealt{torres-flores15}). 
Even if tidal structures are also taken into consideration, only a few studies have shown the flattening in the metal distribution at large galactic radii, such as in the Mice system \citep{Chien07}, NGC 92 \citep{torres-flores14} and NGC 6845 \citep{2015olave}.

In general, all these observational results have been reproduced by simulations (\citealt{rupke10a}, \citealt{perez11}, \citealt{torrey12}), 
which, in some extreme interacting cases result in the inversion of the gas metallicity gradients. The origin of these inverted gradients should be linked with gas inflows produced by the interaction events, as in the case of flat metal distributions (e.g \citealt{werk11}). On the other hand, a few authors have found metallicity drops in the inner regions of dwarf star-forming galaxies. In these cases, metal-poor gas accretion from the cosmic web can explain the metallicity inhomogeneities (e.g. \citealt{sanchez14}, \citeyear{sanchez15}). Such a process may explain the inversion in the metallicity gradient also of high-$z$ galaxies \citep{cresci10}. 
Despite all possible scenarios for the metal mixing process in interacting systems, their observational confirmation in merging galaxies has been severely restricted by the lack of adequate instrumentation. To confirm the nature of the merger events, a bi-dimensional (2D) mapping of the whole system is often necessary, such that both the metallicity and the kinematic of the gas can be probed in detail. However, this scenario is now changing owing to the development of large field-of-view integral field instruments, such as MUSE, which allows an unprecedented high-resolution view of nearby galaxies. In this work, we use MUSE observations of the local interacting/merging system NGC~1487 to map the gas metallicity and kinematic distribution of all the galaxies in the system to probe the nature of its metallicity gradient and to determine the most plausible scenario for its formation.

NGC~1487 has been described as a peculiar Scd spiral, with extended tidal tails and other clear signs of an ongoing gas-rich merger event \citep{aguero97, Fritze}. In addition, its complex morphology resembles other well-known mergers, e.g, the Antennae, the Mice \citep{Wild}, Haro 11 \citep{ostlin15}. In its centre, NGC~1487 displays four bright condensations, firstly described by \citet{aguero97}.
Previous spectroscopic data for some regions of the galaxy display intense emission lines, including some Wolf-Rayet features \citep{Raimann1,Raimann2}. NGC 1487 has a total mass of $10^{10}$ M$_{\odot}$ \citep{aguero97} and \cite{Lee}, studying star clusters in the system, proposed that the age of the merger is $\sim$500 Myrs. Moreover, studies suggest that NGC 1487 is an early stage merger and reveals two clear tails and two nuclei \citep{aguero97,Lee,Georgakakis}. However, the galaxy is in a sufficiently advanced stage to have only one galactic body \citep{Bergvall95}. It is slightly more dynamically evolved (degree of coalescence between the merging galaxies) than NGC 4676 (“the Mice”), but less than NGC 4038/9 (“the Antennae”). 

This paper is organized as follows. In \textsection \ref{sec:data} we describe the observations and the data reduction process. In \textsection \ref{sec:ionizedgas} we study the ionized gas emission using the MUSE data. In \textsection \ref{stellar_pop} we study the stellar populations and mass distribution across the system. In \textsection \ref{sec:kinematics} we study the kinematics of the system. In \textsection \ref{sec:discussion} we promote a discussion about the results and our main conclusions.

Throughout this work, we adopt the distance to NGC 1487 of $D=12.1$ Mpc. We assume a standard $\Lambda$CDM model, with H$_0 = 70.5$ km s$^{-1}$ Mpc$^{-1}$ \citep{Komatsu}, which gives a scale of $0.058$ kpc/arcsec for NGC 1487.

\section{Observations and Data Reduction}
\label{sec:data}

We use archival MUSE data of NGC~1487, which were obtained, reduced and distributed as part of the ``MUSE Atlas of Discs'' \citep[MAD,][]{MAD,MAD1,denBrok} survey. The MUSE spectrograph \citep{2003SPIE.4841.1096H, 2004SPIE.5492.1145B} is an integral field spectrograph mounted at 8m UT4 telescope of the Very Large Telescope of the European Southern Observatory. In particular, observations of NGC~1487 consisted of two datacubes obtained in the Wide Field mode (FoV of 1 arcmin$^2$). These datacubes have spatial sampling of 0.2'' and spectral sampling of 1.25 \r{A}, covering the wavelength range between 4700 to 9300 \r{A}, for one hour on target, with seeing values of 0.702'' and 0.565'' for the two cubes, respectively.

The goal of the MAD is to provide new, detailed view of the local disc population on the small physical scales that are relevant for probing the physical conditions on which gas and stars directly affect each other, and thus the physical origin of the global state of galactic discs.

We use the datacubes reduced by the MAD survey, which uses the MUSE Instrument Pipeline \citep{MUSE} to perform all the tasks of the process, including flat fielding, dispersion and flux calibration, and sky and telluric subtraction. Additional information about the reduction process can be found in \citet{MAD} and \citet{denBrok}. Despite being part of the 45 observed systems in the MAD survey, NGC~1487 is not studied in any of the released papers of the survey \citep{MAD,MAD1,denBrok}. Given the large apparent size of the system, the two datacubes aimed at different locations. For our analysis, we performed the alignment and combination of the cubes using the MUSE Python Data Analysis Framework \citep[MPDAF,][]{MPDAF} code. 

\begin{figure}
    \includegraphics[width=\columnwidth]{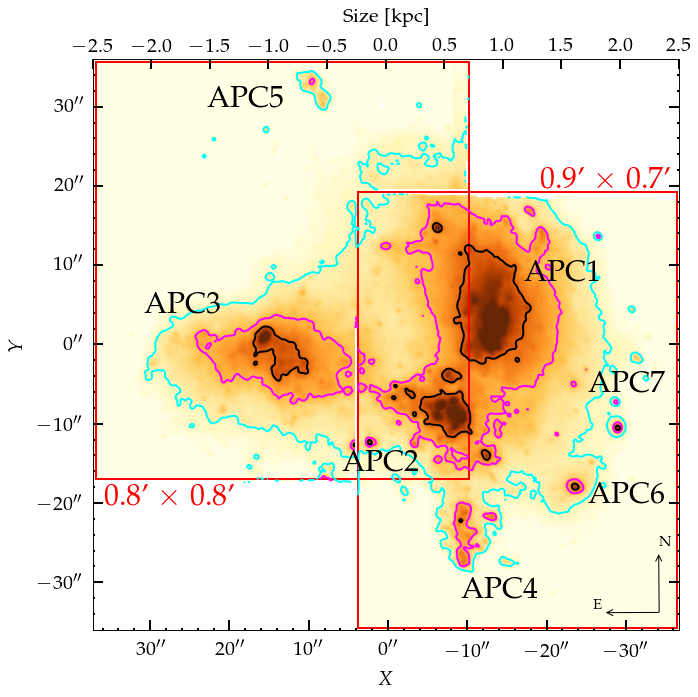}
    \caption[NGC~1487 observed with MUSE/VLT and the definition of the main regions of the galaxy.]{MUSE reconstructed V-band image of NGC~1487. 
    Overlaid are the V-band contours, which delineate the areas where the surface brightness is greater than 18 (black), 19 (magenta) and 20 (cyan) mag arcsec$^{-2}$. Highlighted in darkred are the four condensations present in the merging system NGC~1487, APC1, APC2, APC3 and APC4, firstly proposed by \cite{aguero97} and the tidal tails, APC5, APC6 and APC7, which we propose in this work. White regions within the galaxy indicate masked pixels in the datacube. As red rectangles, we show the MUSE pointings and respective sizes of each cube used in this work.}
    \label{initial}
\end{figure}

For the data analysis, we masked in our datacubes all spaxels detected that may contaminate the measurement of the system. This masking includes low signal-to-noise (SNR < 3) spaxels, saturated stars and its diffraction spikes and all other objects that were not considered part of the system. Masking of these sources was carried out on the white lamp images using the segmentation image produced by {\sc Sextractor} \citep{Bertin} using a $1.5 \sigma$ threshold above the local continuum. 
After the masking, each cube had a size of 0.9' $\times$ 0.7' and 0.8' $\times$ 0.8', respectively. 
The mosaic of MUSE pointings on NGC~1487 is shown in Fig. \ref{initial}, highlighting the four bright condensations proposed by \cite{aguero97}, hereafter APC1, APC2, APC3 and APC4 and the tidal tails proposed by us, APC5, APC6 and APC7.
White regions, including lines inside NGC 1487, indicate masked spaxels.

Overall, the reduced datacubes have very high signal-to-noise ratios (SNR) for all the emission lines we are interested in, so a Voronoi tessellation is not necessary if we are interested only in the gas emission. However, the velocity and velocity dispersion maps are benefited from the use of some binning to increase its SNR. Therefore, to secure a high quality of the analysis, the spectra in each cube were spatially binned in a Voronoi tessellation using the code of \cite{voronoi} to achieve a minimum SNR of $\simeq 30$ pix$^{-1}$ per bin in the continuum. The SNR is measured in the wavelength range $5590 \leq \lambda ($\AA$) \leq 5680$, which does not contain strong emission lines, using the {\textsc der\_SNR} algorithm \citep{2008ASPC..394..505S}. This process resulted in 1087 spatial bins for the two datacubes, with sizes varying from one to $\sim 400$ spaxels, with a median size of 16 pixels. Consequently, assuming approximately circular bins, the typical diameter of each bin is of 0.9 arcsec, or approximately 50 pc in the case of NGC 1487. 

We correct all the spectra for Milky Way dust attenuation assuming a \cite{Fitzpatrick} reddening law with $R_{V}= 3.1$, and assuming a total extinction at the direction of NGC~1487 of $A_{V} = 0.033$, obtained from the \cite{Schlafly} recalibration of the \cite{Schlegel} Galactic dust  map.

\section{Analysis of ionized gas emission}
\label{sec:ionizedgas}

In this section, we focus on the analysis of the emission line diagnostics based on line fluxes and flux ratios, in order to investigate the star formation, ionization and metallicity properties of the ionized gas.

\subsection{Emission line fluxes}
\label{sec:emissionlines}

The emission line maps were obtained using FADO \citep[``Fitting Analysis using Differential Evolution Optimization'', ][]{FADO}, a conceptually novel population spectral synthesis tool with the distinctive property of a) including both stellar and nebular continuum emission in the fit and b) ensuring consistency of the best-fitting star formation history (SFH) with the observed nebular characteristics of a star-forming galaxy (e.g., luminosities and equivalent widths of hydrogen Balmer lines, shape of the continuum around the Balmer and Paschen discontinuity). Besides physical and evolutionary properties for the stellar component (e.g., SFH, mass- and light-weighted age and metallicity), FADO determines emission-line fluxes and equivalent widths (EWs) after correction for underlying stellar absorption.  

\begin{figure*}
\centering
    \includegraphics[width=\textwidth]{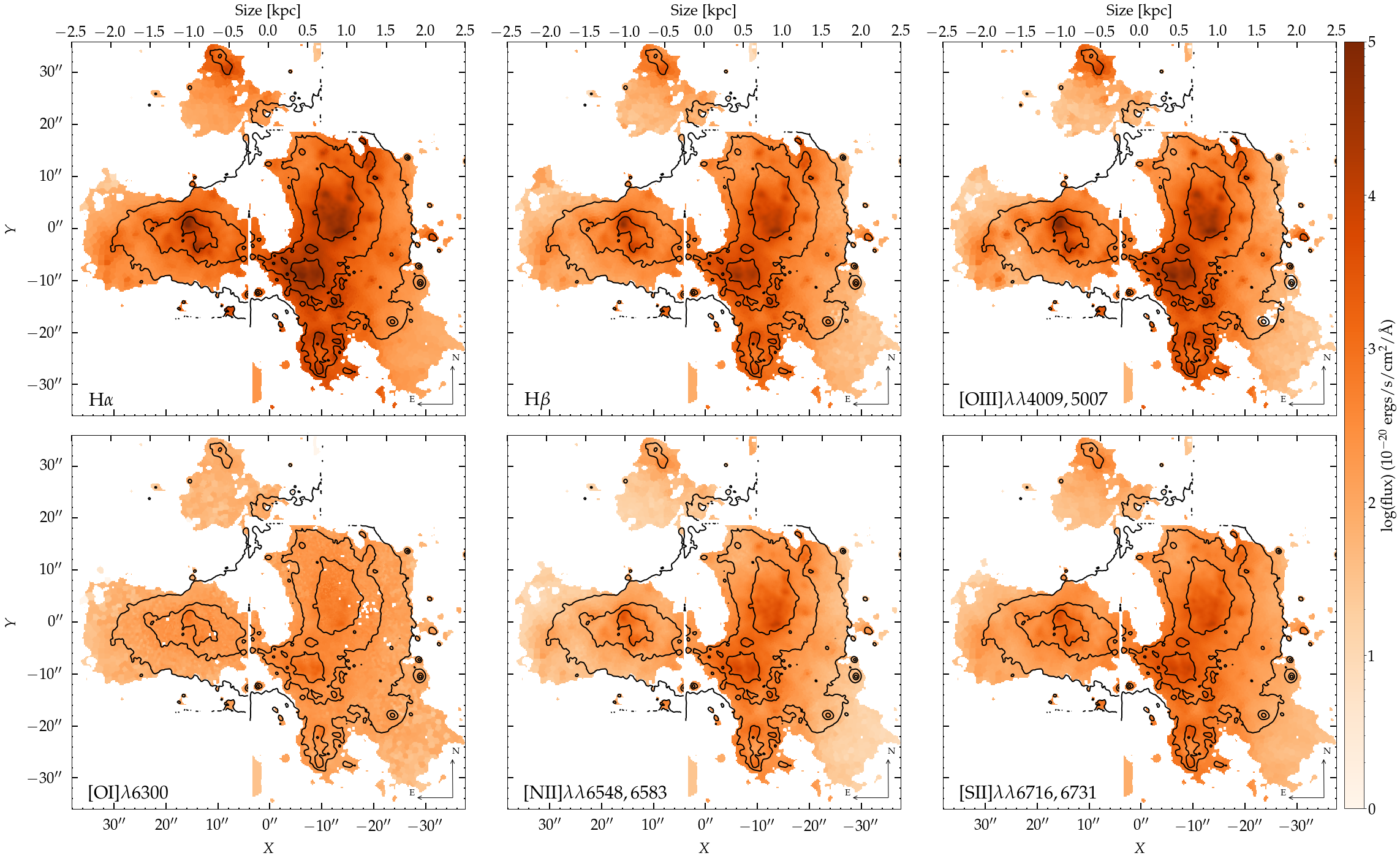}
    \caption[Emission line maps from MUSE dataset of NGC~1487.]{Emission line maps retrieved using the fitting analysis FADO. The panels show the uncorrected for extinction maps of: H$\alpha$, H$\beta$, [OIII]$\lambda5007$, [OI]$\lambda6300$, [NII]$\lambda\lambda 6548, 6583$ and [SII]$\lambda \lambda 6716,6731$, respectively. The black lines represent the V-band isophotes.}
    \label{el_maps}
\end{figure*}

We used FADO in the binned data cubes to determine the flux of six important optical emission lines, including H$\beta$, [OIII]$\lambda4959,5007$, [OI]$\lambda6300$, [NII]$\lambda6548, 6583$, H$\alpha$ and [SII]$\lambda6716, 6731$. 
In Fig. \ref{el_maps} we show the raw emission line flux maps, not yet corrected by extinction, derived from this analysis. 
Interestingly, we can note a high H$\alpha$ emission in the centre of the galaxies, especially in the regions of APC1, APC2, and  APC4.

Furthermore, we take into consideration two known systematic effects that impact the error budget: the errors in the fitting code, and errors related to imperfections in the continuum subtraction. Therefore, to obtain more appropriate uncertainties for our dataset, we use the results from the analysis of \citet{MAD1}\footnote{Notice that \citet{MAD1} uses \texttt{pPXF} \citep{2003MNRAS.342..345C, 2017MNRAS.466..798C} to determine the fluxes instead of FADO, but both provide similar uncertainties according to our tests with both codes. We take into consideration, however, that ``template mismatch'' is an issue that mainly affects pPXF, since this tool uses a limited set of SSPs for fitting the stellar component (and correcting for underlying stellar absorption). A mismatch in stellar age and metallicity will therefore impact stellar absorption corrections, and consequently emission-line profiles (in a manner that is inversely related to emission-line fluxes). Furthermore, unlike FADO, pPXF does not include self-consistent nebular continuum emission, which may impact corrections for stellar absorption in the case of high-sSFR galaxies like NGC 1487.  
The situation is far better in the case of true spectral synthesis, like in FADO or Starlight, where the stellar SED is approximated by a linear superposition of SSPs of different ages and metallicities (up to 300 in the case of Starlight and up to 2000 in the case of FADO), whereby the best-fitting synthetic stellar SED is convolved with a Gaussian convolution kernel to match the absorption-line profile.}.
For the first systematic problem, they used simulations to show that errors in the fitting code for spectra with SNR$ > 3$, such as those that we use in our analysis, is $\sim 2$\%. For the latter effect, they have devised simulations where the continuum subtraction is affected by template mismatch effects, resulting in an additional uncertainty of the order of $\sim 4$\% for the Balmer lines. In the case of NGC 1487, the EW(H$\alpha$) ranges between $\sim 20$ \AA{} and $\sim 200$ \AA{}, thus a possible error of EW $\sim 1$ \AA{} in the underlying stellar absorption profile will introduce a minor error (<5\% in all cases) in H$\alpha$ emission-line fluxes. Therefore, we use the uncertainties provided by FADO conservatively adding a 6\% correction for the systematic error in the line fluxes, but we note that empirical estimates by \cite{MAD1} are based on pPXF, that does not take nebular continuum into account, therefore it is per se non-optimal for the study of strongly star-forming galaxies. That said, empirical error estimates based on pPXF represent rather an upper limit to the true line flux uncertainties from FADO. 

\begin{figure}
    \includegraphics[width=\columnwidth]{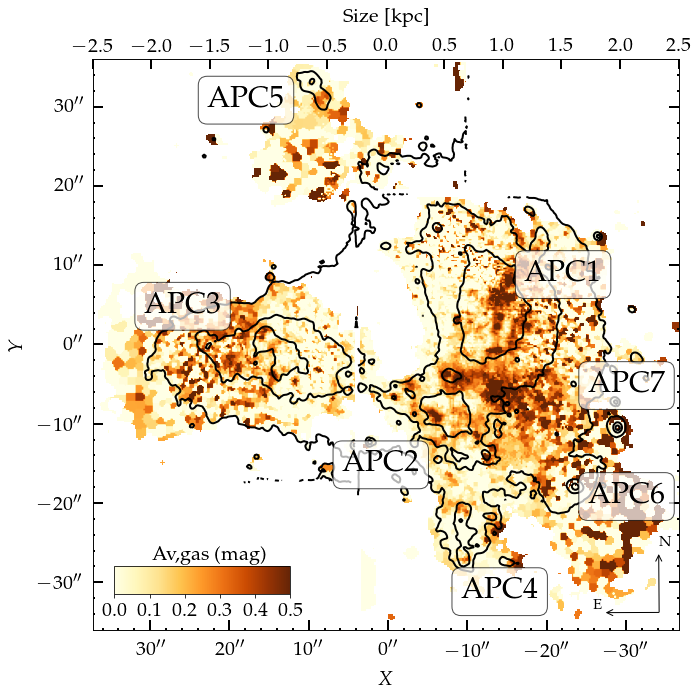}
    \caption{Total dust absorption in NGC 1487, obtained from Balmer decrement. Highlighted in black are the four condensations present in the merging system NGC 1487, APC1, APC2, APC3 and APC4, firstly proposed by \protect\cite{aguero97} and the tidal tails, APC5, APC6 and APC7, which we propose in this work.}
    \label{fig:Av}
\end{figure}

For the correction of the internal dust attenuation, we use the difference between the observed Balmer lines in relation to the theoretical expectations, using the extinction law of \citet{Calzetti} and assuming a case B recombination, with an electron density of \hspace{0.05cm} $n_e = 100$ cm$^{-3}$ and an electron temperature of \hspace{0.05cm} $T_e = 10^4$ K, which gives the predicted ratio (unaffected by reddening or absorption) of H$\alpha /$H$\beta = 2.86$ \citep{Osterbrock}. Fig. \ref{fig:Av} indicates the resulting total dust absorption for our observations, which are used to correct the emission line fluxes of all the lines in this work.

We measure a maximum effective attenuation of $0.92 \pm 0.12$ mag in the transitional from APC1 to APC2 and in the region of APC7. This value is in agreement with the attenuation found by \cite{Mengel} for NGC~1487, and it is typical of star-forming galaxies. This is intrinsically related to the amount of star formation in the galaxy and, for rich-starbursts, it can show extreme values, such as the Mice system \citep{Wild}, which shows a maximum effective attenuation of 7 magnitudes. High attenuation regions in NGC~1487 are observed in the tidal tails and outer-regions of the galaxy as well. 
After correcting for the dust attenuation, we measure a total H$\alpha$ luminosity of L(H$\alpha) = (4.71 \pm 0.42) \times 10^{40}$ ergs/s. From Fig. \ref{el_maps}, we see that the ionized hydrogen emission is concentrated in the central regions of the galaxies and show very low fluxes in the outer-regions and in the NE region of APC1. It is interesting to notice the variation in the H$\alpha$ emission in the tidal tail APC5, where we see a clear transition from a mid-luminosity in the upper region, to very low luminosities below, indicating a possible star forming event in this tail.

\subsection{Determination of the ionization sources from diagnostic diagrams}

\begin{figure*}
    \includegraphics[width=\textwidth]{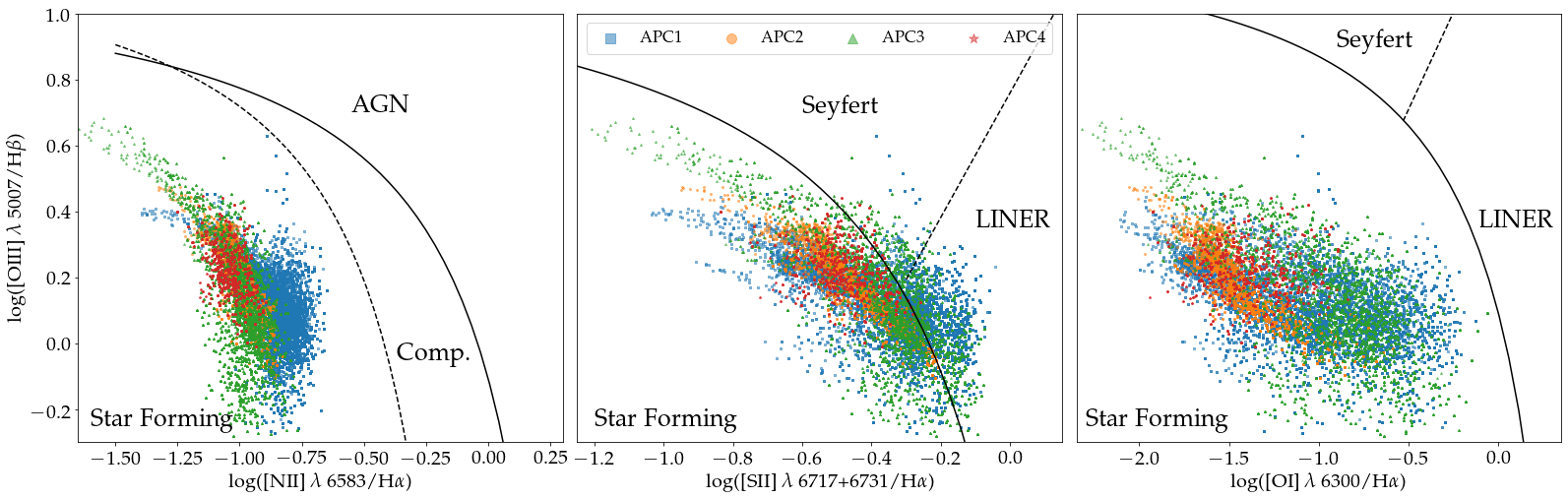}
    \caption{BPT diagrams for the four main condensations of NGC1487, APC1 (blue), APC2 (orange), APC3 (green) and APC4 (red) including BPT-NII (left), BPT-SII (centre) and BPT-OI (right). Left panel: dashed line shows the \protect\cite{Kauffmann} separation between star-forming and composite ionization. The \protect\cite{K06} classification is shown as the solid line. Mid and right panels: solid lines separate star-forming galaxies from active galaxies. Dashed lines separate Seyferts from LINERs.}
    \label{bpt}
\end{figure*}

In Fig.~\ref{bpt}, we show the three classical diagnostic diagrams proposed by \citet{baldwin}, also known as BPT diagrams, to understand the ionization mechanisms of the gas in NGC 1487, indicating the ratio of [OIII]$\lambda 5007$ / H$\beta$ as a function of [NII]$\lambda \lambda 6548,6583/$H$\alpha$ (BPT-NII), [SII]$\lambda \lambda 6717,6731/$H$\alpha$ (BPT-SII) and [OI]$\lambda6300/$H$\alpha$ (BPT-OI). Points of different colours indicate the four regions of NGC 1487, APC1, APC2, APC3 and APC4. As usual, we also show lines based on photoionisation models to distinguish between different ionization mechanisms, including star-forming regions against active galactic nucleus (AGNs) from \citet{Kewley01}, low-ionization nuclear emission-line regions (LINERS) against Seyferts \citep{Stasinska} and star-forming against AGNs  \citep{Kauffmann}.

\begin{figure}
    \includegraphics[width=\columnwidth]{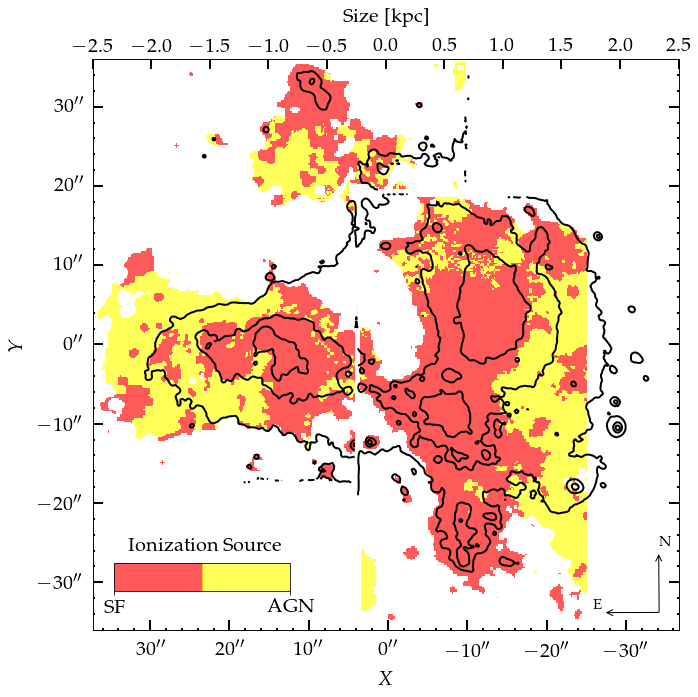}
    \caption{2D BPT SII diagram. The colormap shows the spaxels mainly ionized by star formation (in red) and by AGN/LINER activity (in yellow).}
    \label{fig:2dbpt}
\end{figure}

All BPT diagrams in Fig.~\ref{bpt} indicate that the system is primarily ionized by massive stars. In particular, both BPT-NII and BPT-OI indicate that the ionization levels are well within the limits of the photoionisation mechanisms. However, a few points of APC1 and APC3 in the AGN regions of the BPT-SII diagram indicate a possible influence of LINER ionization too. To investigate the location of these data points, in Fig.~\ref{fig:2dbpt}, we map the classification of the line ratios according to their location in the BPT-SII diagram. In the figure, it is clear that all spaxels associated to AGN/LINER emission are in the outskirts of the condensations, in locations with lower SNR, but not within the central regions of the condensations of the system such as expected for AGNs.

It is observationally established that spatially extended LINER-like emission originates from a variety of physical mechanisms and is present in different environments, such as early-type galaxies (ETGs) \citep{Annibali,porto3d,Gomes}, galaxy bulges \citep[e.g.][]{Ho08,Breda}, diffuse ionized gas (DIG) in intra-spiral arm regions \citep{Singh,Zhang,denBrok} and starburst-driven outflows \citep{Sharp}. In the case of ETGs, \cite{porto3d} conjecture that LINER emission reflects a situation where tenuous gas is exposed to the hard ionizing field from pAGB stars and eventually a Lyman continuum photon leaking AGN. 

For the sake of this work, however, were we are primarily interested in the central condensations of NGC 1487, it seems plausible to assume that the only mechanism involved in the ionization is photoionisation by massive stars.

\subsection{Mapping the electron density}

\begin{figure}
        \includegraphics[width=\columnwidth]{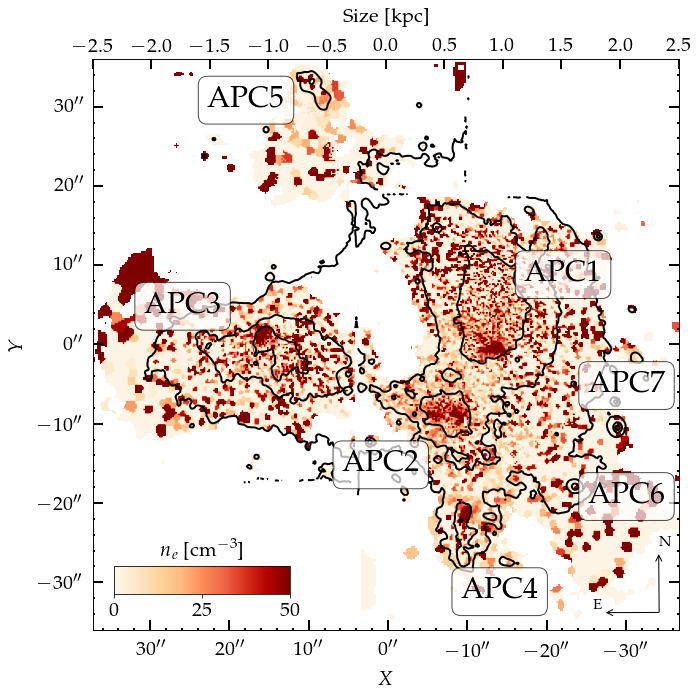}
    \caption{Electron density map of NGC~1487 from the MUSE datacubes obtained from the line ratio [SII]$\lambda6716/\lambda6731$. Highlighted in black are the four condensations present in the merging system NGC~1487, APC1, APC2, APC3 and APC4, firstly proposed by \protect\cite{aguero97} and the tidal tails, APC5, APC6 and APC7, which we propose in this work.}
\label{density}
\end{figure}

In interacting systems, the electron density, $n_e$, can be used to unveil motions and shocks of the gas \citep{Krabbe}. In the wavelength range covered by MUSE, the line ratio [SII]$\lambda6716/\lambda6731$ can be used to estimate the eletron density, where the line ratio decreases with increasing electron density, changing from $\approx1.4$ for electron densities $n_e \approx 100$ cm$^{-3}$ to $\approx 0.4$ for densities $n_e \approx 10^5$ cm$^{-3}$ at $T_e = 10^4$ \citep{Osterbrock}. In Fig.~\ref{density}, we show the electron density map of NGC~1487, obtained from the FADO analysis of the emission lines.

Overall, the electron density is relatively low in the entire system, with relative overdensities only at the centre of APC1 and APC2, and no clear signs of shocks. Even compared to other interacting systems, the measured electron density is relatively small, as interacting systems have electron densities in the range $n_e = 24 - 532$ cm$^{-3}$, typically larger than that of isolated galaxies, which have densities in the range $n_e = 40 - 137$ cm$^{-3}$ \citep{Krabbe}.

\subsection{Gas-phase metallicity and star formation rate surface density}

To determine the gas-phase metallicity of NGC~1487, we use the metallicity-O3N2 relation determined by \citet{MarinoO3N2}, which was derived from integral field observations in the CALIFA survey \citep{2012A&A...538A...8S}, given by 

\begin{equation}
12 + \log(\rm{O/H}) = 8.533[\pm 0.012]\, -\, 0.214[\pm 0.012] \times \rm{O3N2} \mbox{.} 
\end{equation}
According to \citet{MarinoO3N2}, the uncertainty associated to the calibration is $\pm 0.18$ dex ($\pm 1\sigma$), which we also adopt in our analysis.

We calculated the gas-phase metallicities for all spaxels of the cubes, but here we focus only on the parts where the line ratios indicate the gas is primarily photo-ionized by hot stars (APC1, APC2, APC3, APC4 and APC5). 

Flat gradients in interacting systems are expected, but such a flat gradient is expected to only occur in later stages of a merger \citep{Kewley10}. Therefore, studying the metallicity gradient of NGC~1487, we can have an idea of the evolutionary stage of this system. However, knowing that NGC~1487 is already one galactic body, it is expected that the system is experiencing at least some flatness in the gradient currently.

Another indicator of the stage of the merger is the star formation rate of the system. Several studies show that mergers are responsible for creating high star formation episodes. However, \cite{bergvall2003} show that the ignition of the star formation is dependent on the properties of the progenitor galaxies and that a gas-rich merger can evolve to a poor starburst system. 
In the regions of the galaxies where the gas is photo-ionized by light from stars, we can measure the SFR surface density from the dust attenuation corrected H$\alpha$ line luminosity, using the standard conversion to star formation rate \citep{Kennicutt}.

In the regions not dominated by star formation (regions identified to be in the AGN/LINER regions in the BPT diagrams), where the dust attenuation is larger within errors, it is possible that the attenuation correction is leading to incorrect emission line strength estimates and therefore non-reliable SFR surface density estimates. 

\begin{figure*}
    \hspace*{-0.6cm}
    \includegraphics[width=0.83\textwidth]{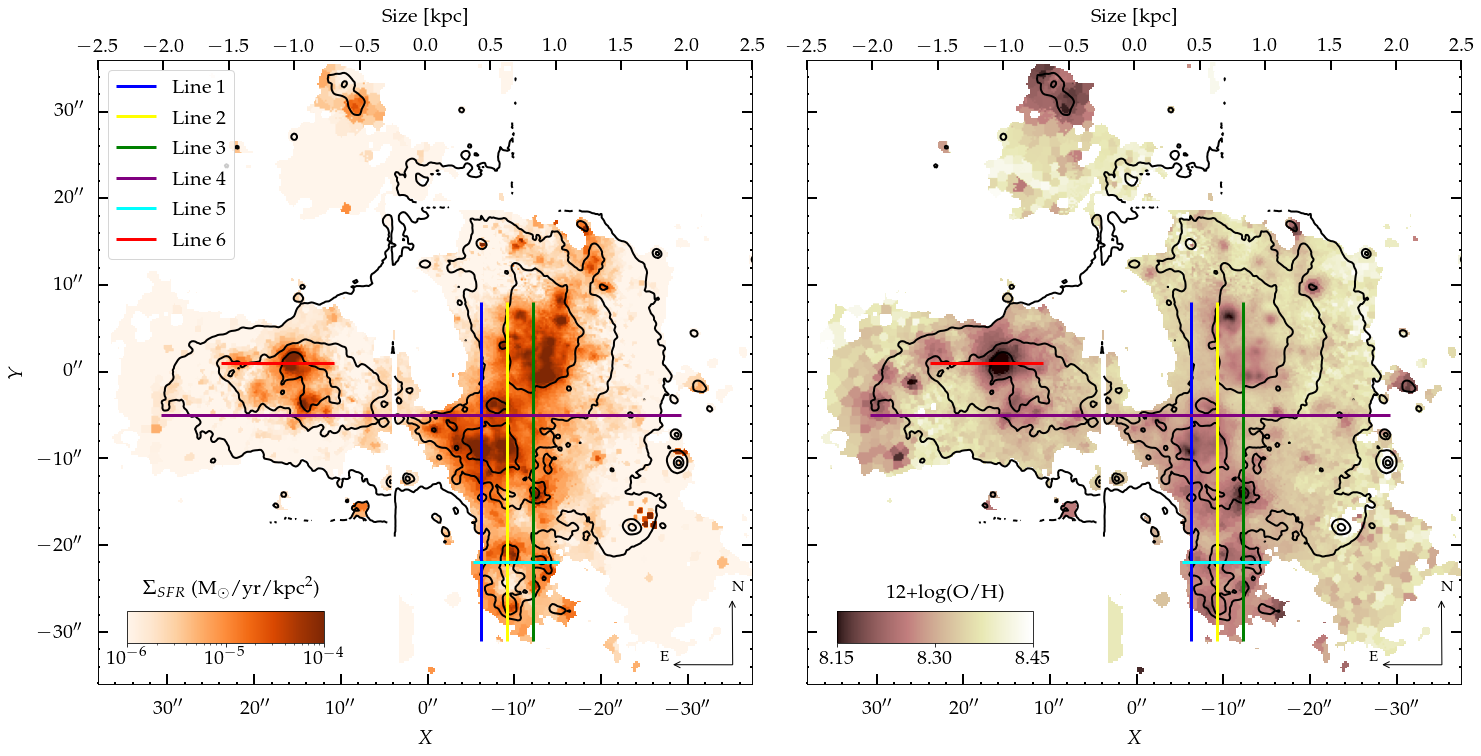}
    \includegraphics[width=0.82\textwidth]{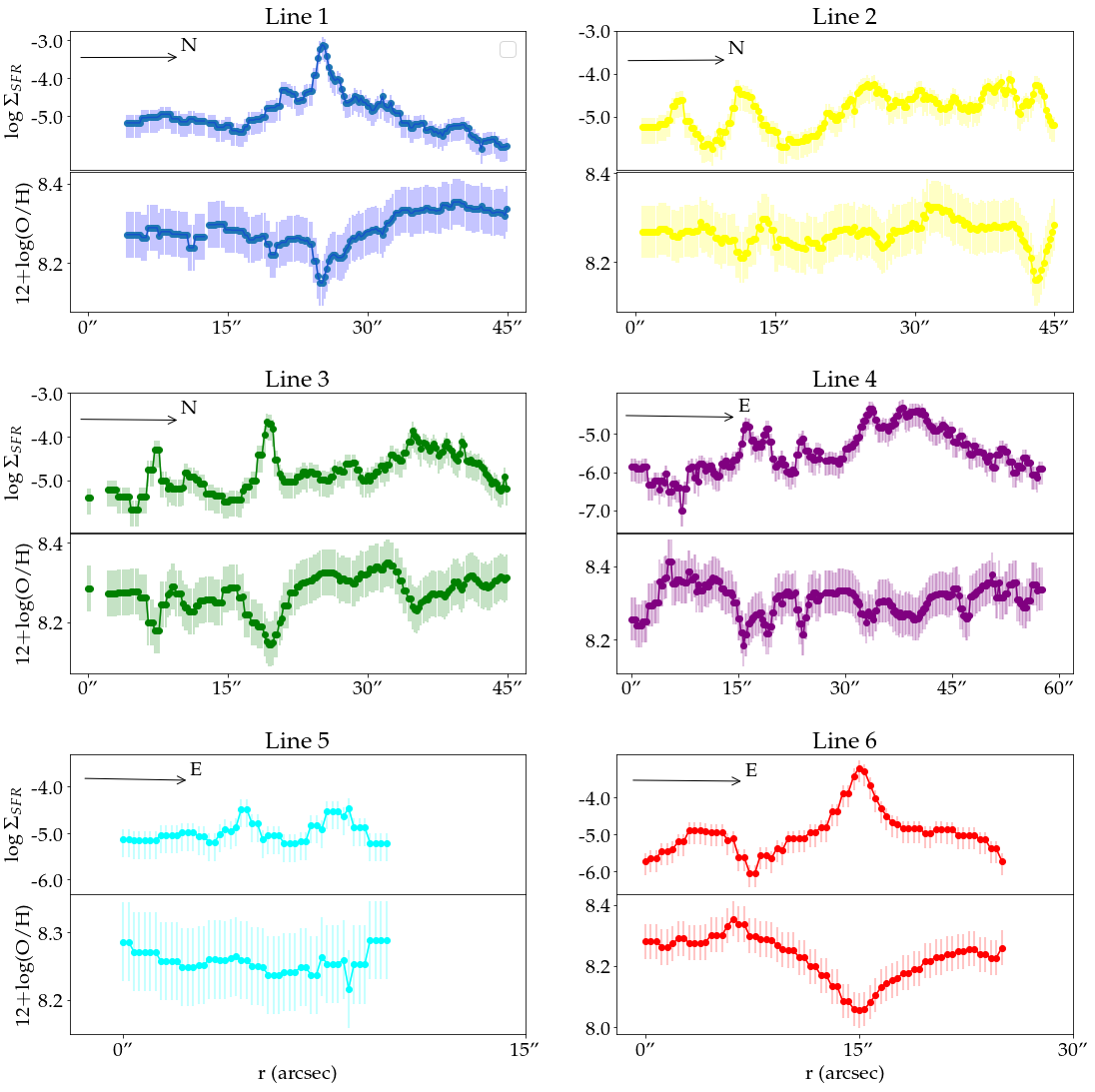}
    \caption{Upper row: star formation rate surface density map in the left and oxygen abundance map of NGC~1487 retrieved using the O3N2 method in the right. The black lines represent the V-band contours constructed from the MUSE data cubes. Each line drawn in the maps represents the displayed gradients shown in the bottom panels. Each gradient plot shows in the top the SFR surface density variation and in the bottom the oxygen abundance variation, with shaded errorbars.}
    \label{met}
\end{figure*}

Fig. \ref{met} shows the oxygen abundance and star formation rate surface density maps retrieved using the O3N2 index and the conversion from \cite{Kennicutt}, respectively.
To trace the metallicity and SFR surface density across the system, we draw lines cutting the system in several positions and directions to guide the estimation of the gradients, these are shown in the bottom panel of Fig. \ref{met}. The errors shown in the bottom panels of Fig. \ref{met} are, as previously mentioned, a combination of the errors due to the calibration/ conversion and to the propagation of the errors on the emission line fluxes.

The metallicity of the gas throughout the galaxies range from $8.3 \pm 0.15$ to $8.8 \pm 0.17$, where the lowest metallicities are found in the four central condensations and in the upper region of APC5, whereas the higher values are in the outskirts of the galaxies. 

In this work we just use the O3N2 calibrator proposed by \cite{MarinoO3N2}. Despite we are using a single method, we note that we are comparing the oxygen distribution across the galaxy, by using the same method. This approach allows us to search for radial trends, independent of the method. In this sense, we note that our O/H measurement does not correspond to absolute values. In order to compare our estimates with other methods, we have estimated the O/H abundance using the method proposed by \cite{Bian}. We found a good agreement between both approaches, consistent with the radial trend previously found from the \cite{MarinoO3N2} calibration.

These findings of an inverted metallicity gradient go in an opposite direction to the results for 38 galaxies within the MAD survey, studied by \cite{MAD1} (none of them being interacting systems). They find a universal trend for negative gas metallicity gradients (radially decreasing metallicity) in the inner regions of the local discs which are sampled in their MAD data. 

We do not attribute the finding of an inverted metallicity gradient for NGC~1487 and to none of the MAD galaxies in the work of \cite{MAD1} as a possible difference between the calibrators used. This conclusion is due to several reasons:
(i) \cite{MAD1} use three calibrators to find the gas-phase metallicity in their work, one of them being the same used by us, i.e \cite{MarinoO3N2}. This means that, even using the same calibrator in the same range of magnitudes ($8 <$ 12+log(O/H) $< 9$), \cite{MAD1} find a negative gradient for the isolated galaxies, while we find a positive one for NGC~1487, an interacting system, so we can only attribute this difference to the physical properties of the galaxy.
(ii) Even using three completely different calibrators in their work, \cite{MAD1} show that they always find the same pattern (a negative metallicity gradient), revealing that even using different methods, we should obtain the same trends, although the metallicity range would change from method to method.
(iii) Even if the calibrators used to obtain metallicities are different in the two works and, therefore, the absolute values may be different, the trends of the data may be similar, meaning that the gradients within a galaxy may be extracted correctly, no matter the method used. This was true for the two methods used in this work, \cite{MarinoO3N2} and \cite{Bian} - note that these two methods show inverted gradient for the system NGC~1487. Assuming that we can generalize this for all methods, we tentatively conclude that a single calibrator on a galaxy should generally allow detection of radial trends on a given object and therefore comparisons between
different objects, even if done with different calibrators, are possible. Having this in mind, the result of no inverted gradient in any of the 38 MAD galaxies, when the gradients were obtained with the method of \cite{MAD1} can be associated with a real difference in the intrinsic physical properties of the galaxies in the MAD sample and NGC~1487. In fact, this can be easily explained by the fact that the latter is a merging object while none of the galaxies studied by \cite{MAD1} were.

As an empirical insight, the areas with lowest oxygen abundances are the ones with the highest star formation rate surface density \citep{sanchez15,SanchezMenguiano, Hwang}, indicating that the merger should be responsible for bringing less-enriched gas to the centre of the galaxies inducing starburst events in these regions. The overall SFR of this galaxy is calculated as $0.37 \pm 0.09 M_\odot/yr^{-1}$, consistent with \cite{Mullan} and \cite{bergvall2003}.

In fact, \cite{bergvall2003} explains why we find so low star formation rates for this seemingly rich starburst system. By studying a sample of 59 interacting/ merging galaxies and 38 isolated galaxies for comparison, using photometric and spectroscopic data, \cite{bergvall2003} find that the interacting and merging galaxies, in comparison to the isolated ones, are more luminous in optical and near-IR and that the dust temperature is higher. However, although this is often claimed as an effect of starburst activity, they favour the explanation that interacting galaxies have higher masses than the non-interacting ones and/or the gas, dust and young stars are more centrally concentrated.
For NGC1487 specifically, they conclude that this is a starburst galaxy with a comparatively low specific SFR, by comparing the emission of the system in the optical (B) and near-IR (H), and although the optical band shows a chaotic signature of a starburst, the near-infrared counterpart outlines the old stellar populations in the galaxy. They claim that this galaxy could be the merger of several dwarf galaxies, instead of two massive ones, what in turn would explain the observed low rates of star formation. 

By analysing the gradients shown in Fig. \ref{met}, considering the uncertainties related to each measurement shaded in the plots, we can see that the central condensations show lower abundances than the outskirts, characterizing an inverted metallicity gradient of NGC~1487. This pattern can be seen specially in the last panel, with the red curve that crosses the central region of APC3 and shows a clear depression in the oxygen abundance in the very central region.

\section{Stellar Populations}
\label{stellar_pop}

\subsection{Stellar mass surface density}
\begin{figure}
    \includegraphics[width=\columnwidth]{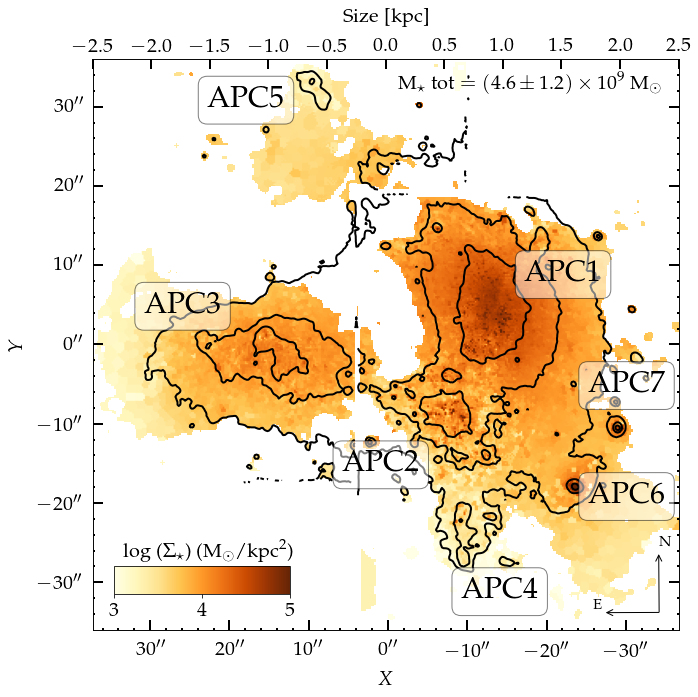}
    \caption[Stellar Mass presently available in NGC~1487]{Stellar mass surface density presently available in the galaxy. The total mass of the system is $(4.6 \pm 1.2) \times 10^{9} M_\odot$, recovered using the Chabrier IMF \citep{Chabrier}. Annotated in black are the four condensations present in the merging system NGC~1487, APC1, APC2, APC3 and APC4, firstly proposed by \cite{aguero97} and the tidal tails, APC5, APC6 and APC7, which we propose in this work.}
    \label{mass}
\end{figure}

Using the best fit spectral energy distribution (SED) from the input simple stellar populations (SSPs) from \cite{BC03}, FADO is able to calculate the most probable stellar mass of the galaxy using the Chabrier initial mass function (IMF) of \citep{Chabrier}.
The map of the surface mass density is shown in Fig. \ref{mass}.

These measurements of surface mass density allow for the calculation of the global mass of the system, estimated to be around M$_{\text{NGC~1487}} = (4.6 \pm 1.2) \times 10^{9}$ M$_{\odot}$, consistent with the results obtained by \cite{aguero97}.
Moreover, from Fig. \ref{mass} we can see that the region of APC1 is a factor of 1-2 dex more massive than the rest of the galaxy, what leads to the conclusion that the dominant stellar body of the galaxy is in this region. 
The tidal tail APC5 has very low masses, both in the upper and lower parts, although it shows a high H$\alpha$ emission and SFR surface density in its upper region.

\subsection{Ages and Stellar Metallicities}
To calculate the ages and metallicities, FADO computes, from the best-fitting population vector, light- and mass-weighted stellar ages and metallicities (cf. \cite{FADO} for details).

\begin{figure*}
    \includegraphics[width=\textwidth]{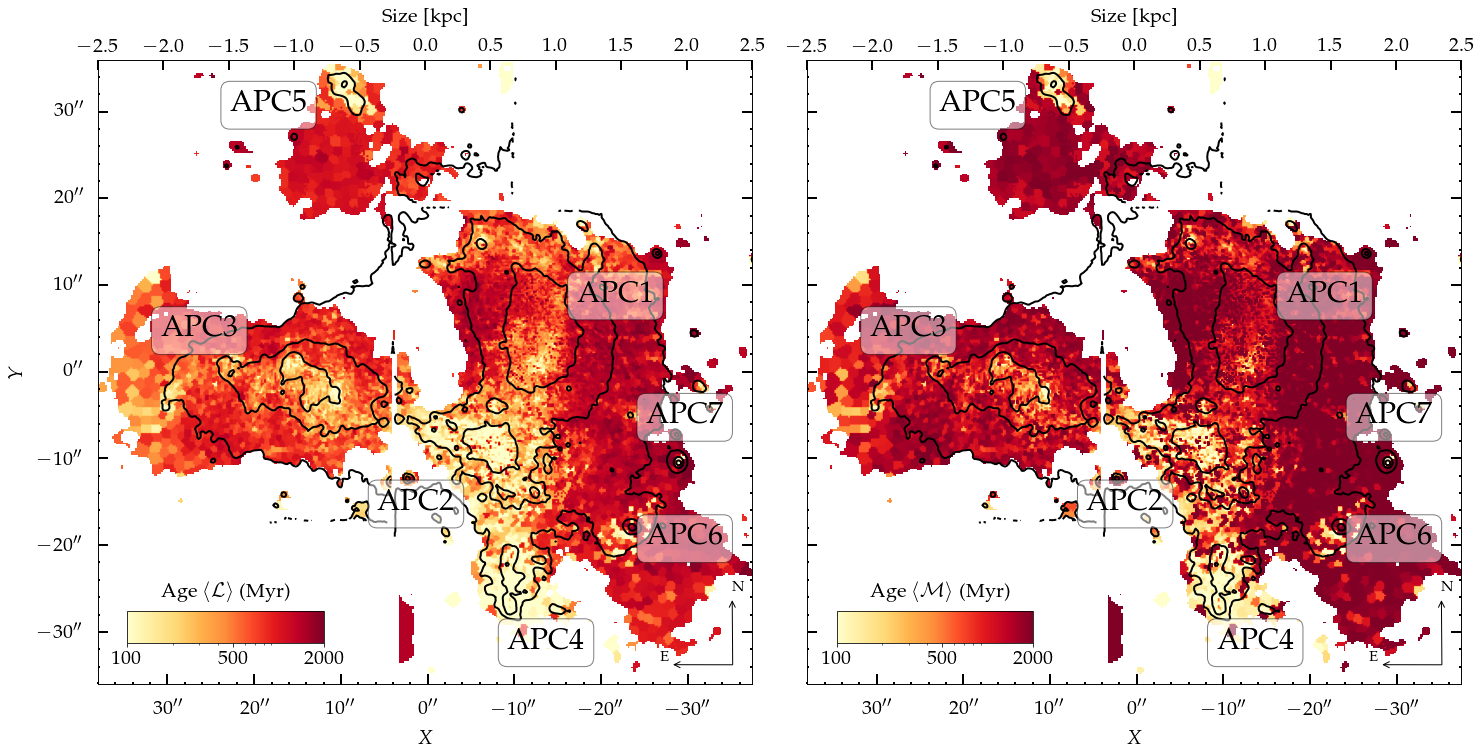}
    \caption[Luminosity and Mass-weighted ages of NGC~1487.]{\textit{Left}, Luminosity-Weighted Age. \textit{Right,} Mass-Weighted Age. Annotated in black  are the four condensations present in the merging system NGC~1487, APC1, APC2, APC3 and APC4, firstly proposed by \cite{aguero97} and the tidal tails, APC5, APC6 and APC7, which we propose in this work.}
    \label{age}
\end{figure*}

The luminosity-weighted stellar age is shown in the left panel of Fig. \ref{age}. This map reflects the most luminous stars within a galaxy, therefore, it tends to highlight younger stellar populations inside the galaxies and their respective ages.

On the other hand, the mass-weighted stellar age, shown in the right panel of Fig. \ref{age}, tends to show the older stellar populations, once the older stars are the ones that hold the majority of the mass inside a galaxy.

\begin{figure*}
    \includegraphics[width=\textwidth]{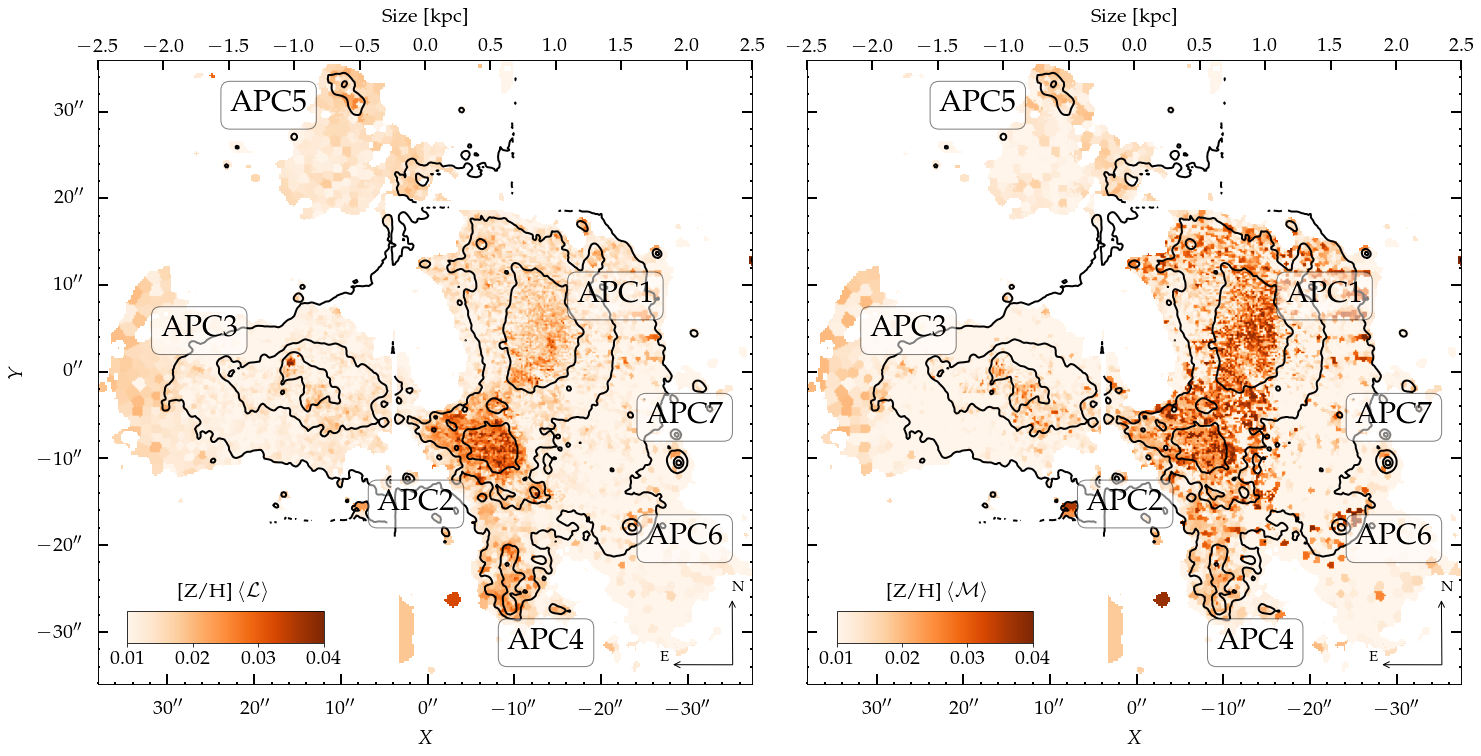}
    \caption[Luminosity and Mass-weighted stellar metallicity of NGC~1487.]{\textit{Left}, Luminosity-Weighted Stellar Metallicity. \textit{Right,} Mass-Weighted Stellar Metallicity both derived using FADO on the MUSE datacubes. Annotated in black rectangles are the four condensations present in the merging system NGC~1487, APC1, APC2, APC3 and APC4, firstly proposed by \cite{aguero97} and the tidal tails, APC5, APC6 and APC7, which we propose in this work.}
    \label{stellarmet}
\end{figure*}

Following the same explanation as before, the light- and mass-weighted stellar metallicity maps are shown in Fig. \ref{stellarmet}, in the left and right panels, respectively.

Converting the observed maps into physical properties, we find an inversion in the age gradient, with the younger stars being found in the centre of the galaxies. The centre of APC2, APC3 and the whole region of APC4 have distinctly younger stellar populations, with light-weighted ages of $\sim100 \pm 8$ Myr or mass-weighted ages of $\sim300 \pm 17$ Myr, consistent with the works of \cite{Lee}, \cite{Mengel} and \cite{Mullan}, which study the star clusters on NGC~1487, specially in the region of APC2. The region of APC1 and outer-regions have the oldest mean stellar ages, as expected since the region of APC1 holds the stellar body of the galaxy, as seen in Fig. \ref{mass}. This region shows ages of about $2.00 \pm 0.34$ Gyr (light-weighted) or $3.00 \pm 0.42$ Gyr (mass-weighted). Here, it is important to reinforce that we do not rule out an older stellar population to the system. The presentation of the ages between 0 and 2 Gyr is just to enhance the contrast on Fig. \ref{age}. Actually, it is apparent from the SFHs in Fig. \ref{sfh}, that an old stellar component of $\sim$ 10 Gyr is present in regions APC1 and APC3 (also a stellar population of $\sim 4$ Gyr in APC2).

We can see that both the light- and mass-weighted maps showing the ages and metallicities of the stellar populations are very similar, since they reflect the same stellar populations. As we can see for NGC 1487, low metallicity stars are older, and higher metallicities are attributed to younger stellar populations, which are born in the metal-enriched by past populations interstellar medium.

It is interesting to compare the stellar metallicity map with the gas-phase metallicity one, shown in Fig. \ref{met}, as they show opposite trends. As we discussed previously, it is usual to see inverted oxygen abundance gradients in interacting systems, which we observe in NGC~1487. For the stellar metallicities, however, the most metal-enriched regions are the centre of APC1 and APC2, both in the mass- and light-weighted maps, and the less-enriched areas are in the outer-regions of the galaxy. In the case of NGC 1487, we see that ongoing SF is spatially confined to the nuclei of the interacting counterparts. These nuclei, which presumably are chemically enriched, most likely still dominate the light-weighted metallicity, even though young luminous stars forming out of slightly less enriched gas are present (``slightly'' because the 12+log(O/H) at the peak of SFR in lines 1, 3 and 6 (in Fig. \ref{met}) is just by $\sim 0.1$ dex lower than in the surroundings). Also, if SF in the nuclei is relatively recent (20-30 Myr), as we see in Fig. \ref{sfh}, then the nucleosynthetic products from the newborn stars (i.e. the hot metal-enriched plasma produced by type-II supernovae) will not have enough time to cool down and become detectable through optical spectroscopy.

\begin{figure*}
    \includegraphics[width=\textwidth]{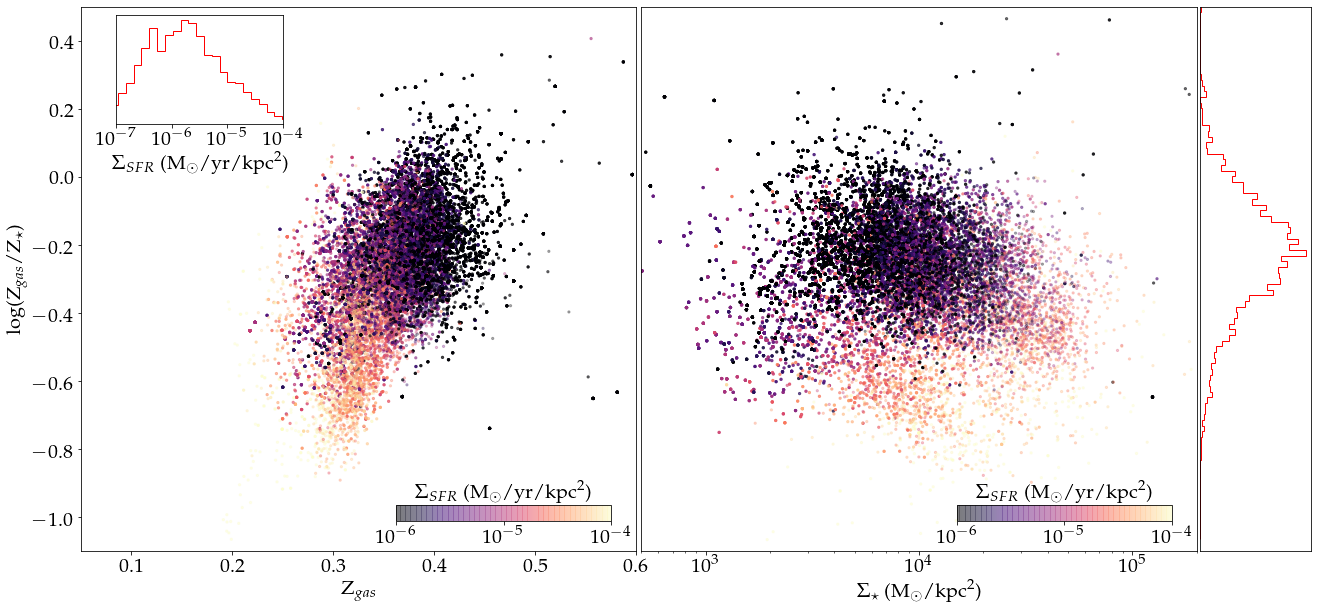}
    \includegraphics[width=\columnwidth]{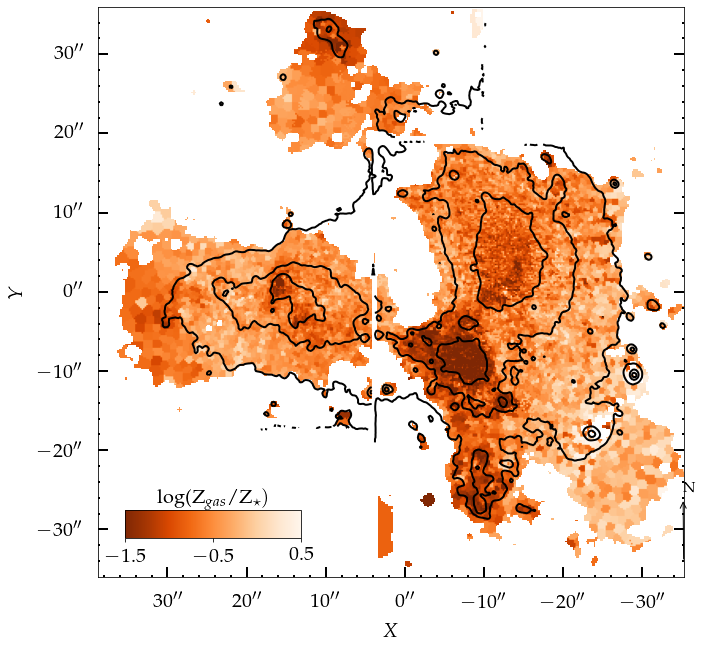}
    \caption{Upper left, gas-phase metallicity vs. relative abundance of gas-phase and stellar metallicities color coded with the SFR surface density, the inset histogram highlights the distribution of the SFR surface density for both scatter plots in the upper row. Upper right, surface mass density vs. relative abundance of gas-phase and stellar metallicities colored with the SFR surface density, the histogram at the right hand side of the scatter plot highlight the marginal distribution of log(Zgas/Z$\star$). Bottom, map of the relative abundance of gas-phase and stellar metallicities throughout the system.}
    \label{met_all}
\end{figure*}

In order to compare the gas-phase and stellar metallicities, we produced the three panels present in Fig. \ref{met_all}, where we show: (i) the variation of the gas-phase metallicity with the relative gas and stellar metallicities, colored with the SFR surface density. (ii) the variation of the surface mass density with the relative gas and stellar metallicities, colored with the SFR surface density. (iii) the map of the relative gas-phase and stellar metallicities. 

The upper left and upper right panels of Fig. \ref{met_all} tell us that regions with the highest SFR surface density ($\Sigma_{SFR}$) or mass surface density ($\Sigma_{\star}$) (i.e. the nuclei of the merging counterparts) have the lowest gas-phase metallicities - both in absolute terms and relative to the chemical abundance of the stellar component, while the bottom panel of Fig. \ref{met_all} highlights the relative abundance of gas and stellar metallicities in the system and clearly shows where these gas under-abundances are most pronounced, i.e. APC2 and APC4.

From Fig. \ref{met_all}, we can see that there is a tendency for the warm gas that is associated with the regions showing the highest $\Sigma_{SFR}$ to show a lower metallicity than the more diffuse/lower-surface brightness gas by a factor of $\sim$ -0.3 dex.
A possible interpretation for this is that this high-$\Sigma_{SFR}$ gas component
originates from the less metal-enriched periphery of the merging counterparts and have not had enough time to become self-enriched by the current SF episode.

\section{Kinematics}
\label{sec:kinematics}
The kinematics of the system was extracted using the penalized pixel-fitting (pPXF) code \citep{ppxf}, which consists of modelling the observations by a combination of spectral templates convolved with a line-of-sight velocity distribution, parametrized as a Gauss-Hermite profile \citep[see][]{gerhard}. pPXF is thoroughly presented in \cite{ppxf}. 

\begin{figure}
    \includegraphics[width=\columnwidth]{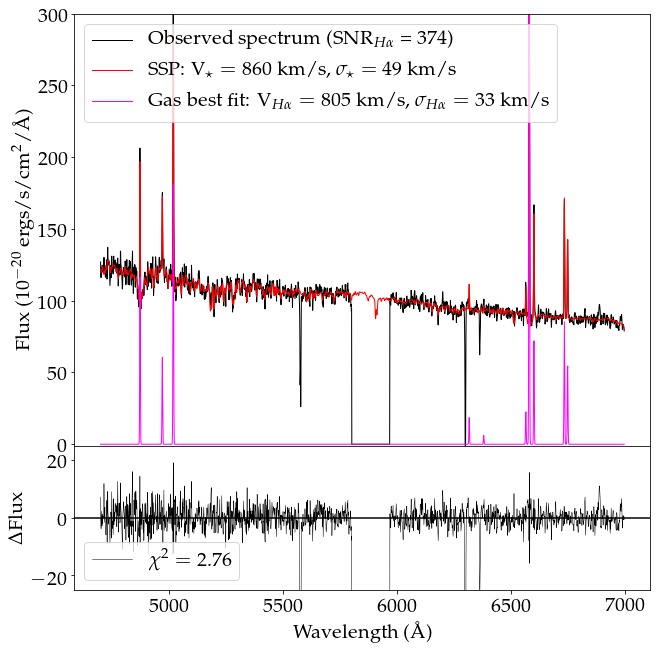}
    \caption{Example of the fitting process with ppxf for the MUSE observations of NGC~1487. Upper panel: The observed spectrum is shown by a black line, the best fit for the stellar component is shown in red, and the best fit for the emission lines is shown in pink. Bottom panel: black line shows the residuals of the fit.}
    \label{fig:example_ppxf}
\end{figure}

We fitted each spectrum with eight kinematic components, one stellar and seven focused on specific nebular emission lines (H$\alpha$, H$\beta$, [OIII]$\lambda 5007$, [OI]$\lambda 6300$, [SII]$\lambda 6716$, [SII]$\lambda 6731$ and [NII]$\lambda \lambda 6548, 6583)$. The gas velocities are computed using a set of Gaussian emission line templates from the EMILES library \citep{emiles}, while the stellar velocities are calculated using Simple Stellar Population (SSP) templates from \cite{Vazdekis}, constructed with stellar spectra of the EMILES library.
In Fig. \ref{fig:example_ppxf}, we show an example of the fitting process of pPXF for one of the bins of NGC~1487, after the Voronoi tesselation is performed.

\begin{figure*}
    \includegraphics[width=0.67\textwidth]{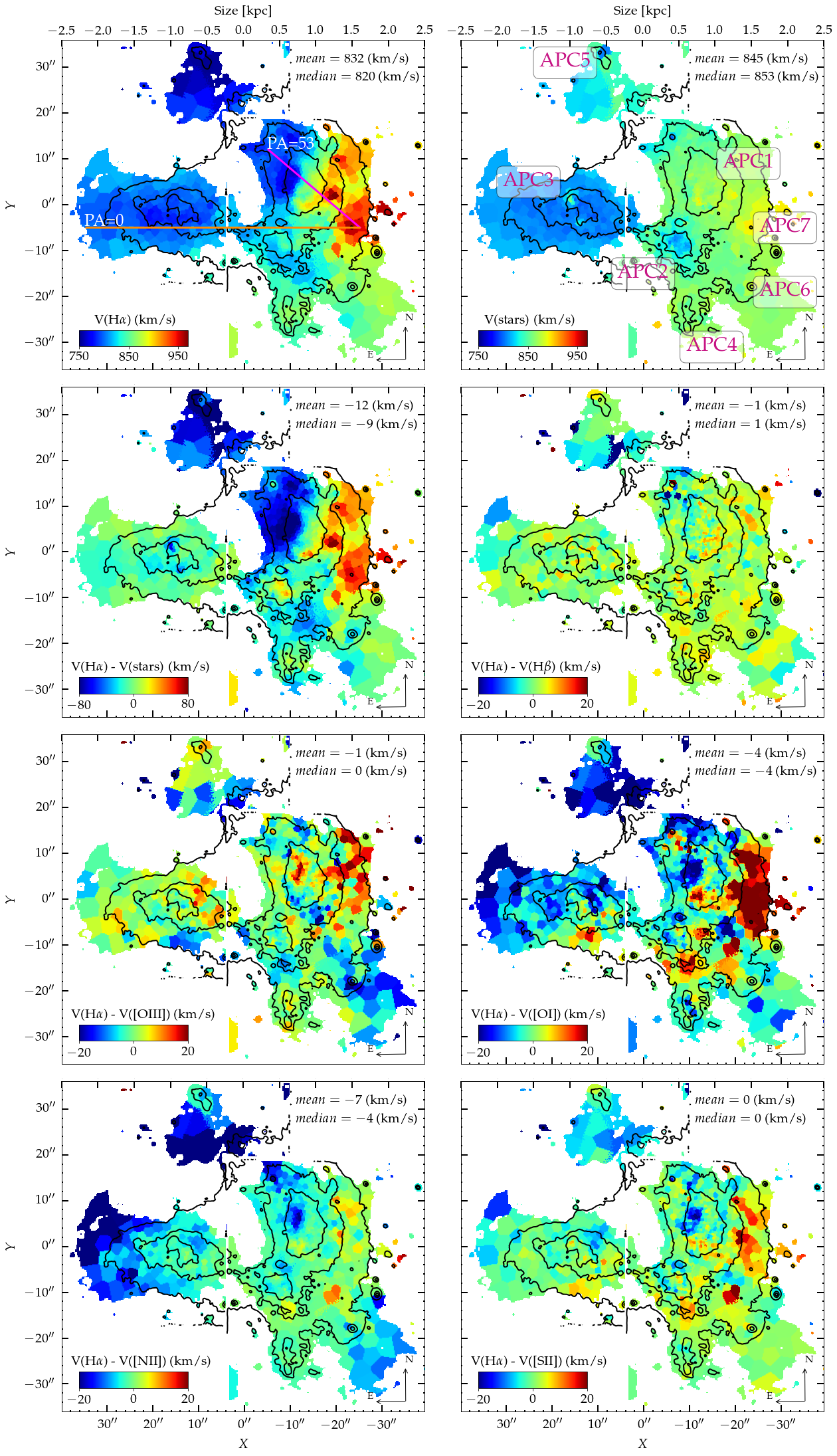}
    \caption{Velocity fields of NGC~1487 obtained using pPXF with seven nebular and one stellar component. The first row shows the velocity of the H$\alpha$ line and of the stars, and the following panels show the velocity of the stars and of the other emission lines compared to H$\alpha$. Respectively: stars, H$\beta$, [OIII]$\lambda 5007$, [OI]$\lambda 6300$, [NII]$\lambda \lambda 6548, 6583$ and [SII]$\lambda\lambda 6717,6731$. In the first panel, the two position angles assumed are highlighted: in magenta (PA=53 deg) shows the angle of the rotating pattern of APC1, and the golden line (PA=0) is the angle assumed of the system as a whole. In the second panel, annotated in magenta, are the four condensations present in the merging system NGC~1487, APC1, APC2, APC3 and APC4, firstly proposed by \protect\cite{aguero97} and the tidal tails, APC5, APC6 and APC7, which we propose in this work.}
    \label{fig:velocity}
\end{figure*}

\begin{figure*}
    \includegraphics[width=0.7\textwidth]{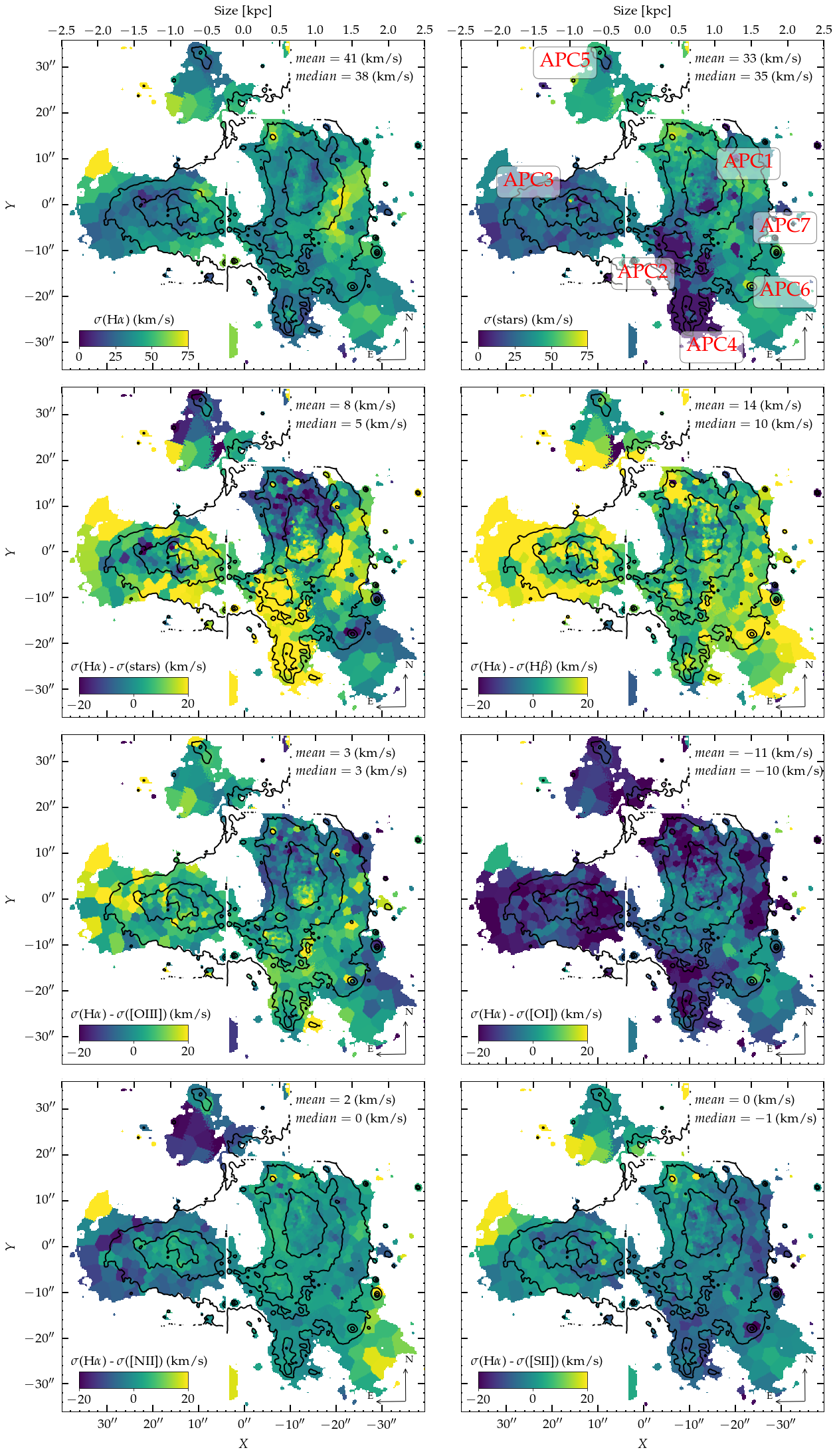}
    \caption{Velocity dispersion maps of NGC~1487 obtained using pPXF with seven nebular and one stellar component.  The first row shows the velocity of the H$\alpha$ line and of the stars, and the following panels show the velocity of the stars and of the other emission lines compared to H$\alpha$. Respectively: stars, H$\beta$, [OIII]$\lambda 5007$, [OI]$\lambda 6300$, [NII]$\lambda \lambda 6548, 6583$ and [SII]$\lambda\lambda 6717,6731$. In the second panel, annotated in red, are the four condensations present in the merging system NGC~1487, APC1, APC2, APC3 and APC4, firstly proposed by \protect\cite{aguero97} and the tidal tails, APC5, APC6 and APC7, which we propose in this work.}
    \label{fig:dispersion}
\end{figure*}

With this analysis, we recovered the velocity field and velocity dispersion maps of each emission line and of the stellar component. The velocity field provides important information regarding the gravitational support of NGC~1487, while the dispersion map is useful to understand the turbulence in this system.  
Figs. \ref{fig:velocity} and \ref{fig:dispersion} show the resulting maps of the LOS velocity and velocity dispersion of NGC~1487, for each emission line.
In these maps, we show, in the first row, the velocities of the H$\alpha$ line and stellar LOS velocities, respectively. In the following rows, we show the velocities of the other lines with respect to H$\alpha$.
Typical uncertainties in the measurements of the LOS velocities are of the order of $7$ km s$^{-1}$, calculated as the mean uncertainty considering all the data.

To double check our results, we have also derived the velocity fields and velocity dispersion maps using both {\sc FADO} and {\sc PORTO3D} \citep{porto3d,Gomes}, which uses {\sc STARLIGHT} \citep{Starlight} for fitting the stellar spectrum and incorporate a suite of routines for measuring emission-line fluxes and kinematics. All of the methods recovered the same patterns as shown in Figs. \ref{fig:velocity} and \ref{fig:dispersion}, with pPXF. 

The stellar velocity field shows two distinct velocity regimes: the eastern region, APC3, that displays a constant velocity of 812$\pm8$ km/s and the western region (APC1-2-4-6) that also shows a constant but higher velocity of $862\pm14$ km/s.  The northern region APC5 shows an intermediate velocity of $834\pm 11$ km/s. This East-West gradient of $\sim$50 km/s could indicate a global rotation of the system as a whole, around a south-north axis on a 4 kpc-scale.

The gaseous velocity field shows globally a similar velocity gradient to the stellar component (assuming a PA $\sim$ 0), of $\sim 50$ km/s, but in addition to this large-scale velocity gradient, the gas displays a more complex local behaviour, within a region embedding regions APC1 and APC7. In particular, a velocity gradient as large as $\sim 200$ km/s is observed over a scale of $1.6$ kpc, aligned along a PA of $\sim$53 degrees, while the stellar velocity gradient is only $\sim$40 km/s over the same area.

Over the whole system, the mean and median radial velocity differences between the gaseous and the stellar components are low: $-12$ and $-9$ km/s respectively.  However, this hides velocity differences that are much larger in two areas: 
1) APC1/APC7 where a velocity difference of $\sim 160$ km/s is observed between both components, over a scale of 1.6 kpc and
2) The gas component in region APC5 shows a radial velocity $60\pm20$ km/s lower than the stellar component. In addition, small velocity gradients of $\sim 64$ and $\sim 30$ km/s, respectively, are observed in both components, but presenting an almost perpendicular major axis.  
 
In other words, the radial velocities of the gaseous and stellar components roughly match in the southern part of the galaxy (regions APC3-2-4-6) but they disagree in the northern part (regions APC1-5-7), where large differences are seen.  It is also noticeable that even if the mean velocity differences are low, on average, the gaseous component displays lower velocities than the stellar component in regions APC5 ($-60\pm20$ km/s), APC1-East ($-55\pm21$ km/s), APC3 ($-7\pm10$ km/s) and higher velocities in regions APC1-West ($+37\pm20$ km/s) and in the brightest central region of APC2 ($+15\pm15$ km/s).

The irregular velocity fields displayed in both stars and gas maps as well as the difference between the velocity patterns of those two components suggest to us that one possible scenario may be that this system is a dwarf irregular galaxy undergoing a merging process.  

Typically 90\% of the mass of a late-type disc is in the stars and only 10\% in the gaseous component, so the potential well is mainly traced by the stellar masses (if we exclude dark matter, of course). In addition, stars are collisionless, whereas the gas is collisional, meaning that the collision between clouds create shocks, sound waves, star formation, turbulence, dissipate energy by different processes (including supernovae explosion, etc...) and thus, the gas is not relaxed during the merging phases (until the gas lies again in a disc, which is usually quite fast), and this is exactly why we can see that the galaxies underwent collisions because of the unrelaxed status of the dissipative component. 

Furthermore, the observed velocity difference between the gas and the stars in the northern part of the galaxy provides evidence for gas accretion coming from the north. The fresh gas accreted around APC1, the most massive structure of the system, is falling into the potential well, which induces the observed rotation.  This fresh gas is also accreted around the second brightest clump APC2, where higher velocities are observed for the gas than for the stars.

An alternative scenario to a single dwarf irregular galaxy described above is the merging of several clumps (two major ones, APC1+APC2 and APC3, and other smaller ones - the remaining clumps). In this second scenario, APC1 could be a giant expanding gaseous bubble (considering its size of $\sim$0.5 kpc), resulting from intense star formation, and in that case, this bubble would overlap a larger region including APC1 and APC2, the two main components of the system, displaying a constant global radial velocity of $\sim860\pm 10$ km/s, i.e. with no rotation.  In both scenarios the stellar and the gaseous velocity components do not match, favouring a merging scenario, in either case.\\
In a scenario that includes the merging of different structures, regions APC1 and APC2, located near the centre of gravity of the system, could be in a more advanced phase of merging while the peripheral regions APC3, APC4, APC5, APC6 and APC7 could be in the pre-merging phase. This would explain the large-scale velocity differences observed, mainly between APC3 and APC1 in the stellar component. In a merging scenario, this 50 km/s velocity difference could come from the difference of systematic velocities between the progenitors, one approaching the other before the merging. This assumption is not in contradiction with an overall rotation of the system because, separated by small distances, it is very likely that gravity has caused these two objects to rotate around one another.\\

The velocity dispersion measurements are mainly affected by the line spread function (LSF) of the MUSE spectro-imager. The displayed velocity dispersion maps are corrected for this instrumental broadening. Some of the corrected gaseous velocity dispersions are below the instrumental resolution. At the wavelength of H$\alpha$, the dispersion of the LSF is $\sim\sigma_{LSF} \approx 47$ km s$^{-1}$, meaning that it is not realistic to consider velocity dispersions smaller than $\sim\sigma_{LSF}/2\simeq 25$ km/s \citep{Boselli}. Velocity dispersions of ionized gas in disc galaxies range from 15 to 50 km/s \citep[e.g. ][]{Epinat,ErrozFerrer}. As the thermal width is typically of the order of 10 km s$^{-1}$, most of the behaviour observed must be due to turbulence. \\

The gas velocity-dispersion maps show that the profiles are overall uniform, with minor changes across the system. Again, we can see that there are almost no differences between the Balmer and forbidden lines, except in the areas with low SNR. As for the stellar velocity dispersion map, we see narrower trends in the centre of APC1, APC2 and APC4 than in the remaining regions. The region with highest stellar dispersion velocities coincide with the region where the majority of the mass of the system is located. This high-dispersion area coincides also with the gas rotating area, shown in the previous figure (Fig. \ref{fig:velocity}).\\
From the combination of the velocity field and of the velocity dispersion map, we conclude that APC1 is undergoing a phase of merging during which the gas, out of equilibrium, shows disordered motions. The duration of this phase is relatively short because the gas is collisional and tends to reform a disc relatively quickly. It is interesting to notice that this apparently rotational pattern is around the centre of mass of the galaxy, shown in Fig. \ref{mass}.\\
In merging or even pre-merging phases, collisions between gas clouds dissipate energy and create turbulence, stars and gas clouds are both subject to strong tidal forces, resulting in disordered motions, slowing down the rotation. We can guess that the undergoing merger of gaseous structure will rebuild an unique rotation body if stellar relaxation processes do not destroy it. This competition is driven by the system gas fraction.\\ 

\section{Discussion and Summary}
\label{sec:discussion}

In this work we studied the merging system NGC~1487, using IFS data from MUSE, together with other ancillary data from the literature. 
In the following we summarize the current evolutionary stage of the system and our main results, given the insights we obtained in the physical and kinematic properties. 

\subsection{Morphology} NGC~1487 is a system with few studies in the literature to date and the past studies have not reached a consensus about the evolutionary stage of the system. \cite{aguero97} suggest that the galaxy is an ongoing merger between two gas-rich galaxies, suggestion followed by \cite{Lee}. On the other hand, \cite{bergvall2003} show, comparing optical and infrared data, that the system is most probably an ongoing merging event between dwarf galaxies in a small group, instead of the merger of two massive galaxies. \cite{Georgakakis} observed NGC~1487 in radio wavelenghts (4.79 GHz (6 cm) and
8.64 GHz (3 cm)), and concluded that the system is in a pre-merger stage, and it is a result of a gas-rich merger, which has not coalesced yet.

In this work, having the physical and kinematic information of the ionized gas and stellar components of the system, we favour the hypothesis that the origin of this interaction is a merger between dwarf galaxies forming a small group. 

Several pieces of evidence indicate that the progenitors of the merger are gas-rich dwarf galaxies:\textit{ (i)} the low mass measured for the system as a whole and for its condensations (see Fig. \ref{mass}, Table \ref{tab:apc}). \textit{(ii)} the small velocity gradients between APC1, APC2 and APC4. The difference between these three condensations from APC3 is larger ($\sim 50$ km s$^{-1}$), but it is still a small difference if compared with the expected variation between two massive galaxies. \textit{(iii)} the physical properties that are not yet homogenized throughout the entire system, i.e. four centres of star formation (APC1, APC2, APC3 and APC4), that show different masses, ages and metallicity distributions. \textit{(iv)} the observed low star formation rates of the system, consistent with the behaviour proposed by \cite{bergvall2003}.  

According to the classification of evolutionary stages of Hickson Compact Groups by \cite{Verdes}, a system is in an intermediate phase when there are multiple tidal tails formed and a large amount of the atomic gas is in the intragroup medium, which is the case for NGC~1487, as shown by \cite{Mullan}. The classification of \cite{Verdes} was used to prove that HCG~31, one of the most striking Hickson Groups, is an ongoing intermediate-phase merger of multiple low-mass objects \citep{Amram}. Considering that our results point to NGC 1487 being the result of a merger of dwarf galaxies, it is interesting to compare it with HCG~31. As we show for NGC~1487, HCG~31 also showed a flat metallicity gradient \citep{torres-flores14} and perturbed velocity fields \citep{Alfaro}. 

Regarding the evolutionary stage of the system, \cite{Surace} classified interacting systems in five dynamical stages, first approach, first contact, pre-merger, merger, and old merger.
We propose that NGC~1487 belongs to the class of pre-merger, described by \cite{Surace} as ``two nuclei with well-developed tidal tails and bridges'' \citep[same classification received by HCG~31, ][]{Amram}. This pre-merger scenario is consistent with the one proposed by \cite{Georgakakis} for NGC~1487.

There are several pieces of evidence that indicate to us that this system may evolve to the merging stage. These are: (1) \cite{Xu} show that close pairs with overlapping discs that have starbursts in their overlapping regions (in this case, the region that connects APC3 to the regions of APC1, APC2 and APC4 that shows the highest SFR surface density of the system) can be either advanced mergers or severely colliding systems. In addition, \cite{Xu} show for one out of the three advanced-merger systems in their sample, that the starburst in the overlapping region is stronger than the central starburst, which is exactly the case for NGC~1487 (APC2 has higher SFR than all the rest of the system, see Table \ref{tab:apc}). These arguments are strong indicators that the galaxies are either severely colliding or merging. (2) The system very clearly has tidal tails. This indicates that the interacting galaxies have had at least one earlier passage, i.e., the galaxies are in a bound orbit \citep{Bournaud2}. (3) The galaxies are now lumped into one single galactic body, although showing distinct velocities and inhomogeneous distributions of physical properties.

As previously mentioned, \cite{Verdes}, using optical and HI data to study a collection of compact groups,  small dense groups of galaxies similar to what is found in NGC~1487, concludes that a  gas-rich system in a pre-merger stage shows two distinct galaxies with well-developed tidal tails and bridges as a result of the first close encounter \citep{Barnes,Mihos,Springel}. Although we do not use HI observations of the system in this work, tidal features are clearly observed within NGC~1487, reinforcing our hypothesis that this system is most probably in this evolutionary stage. Since the early simulations of \cite{Barnes90b,Barnes90} and \cite{Athanassoula}, it is well known that the merging times of small groups is short - all galaxies merge within a few crossing times. Galaxies within groups will firstly form pairs or accrete one another, but these interacting pairs and perturbed galaxies will quickly come closer and closer until they all fall in, to form a resulting giant elliptical galaxy (or a dwarf, in our case, given the mass of the system), with no time to form a mature remnant within the group environment. This scenario was corroborated by previous works focused on the studies of the galaxy populations of compact groups \citep{Claudia, Claudia2005, Proctor2004}, who showed that the elliptical galaxies in these small groups are not the result of mergers, they are, instead, old and metal rich systems, more similar to elliptical galaxies in clusters, suggesting that the merging time of these small groups is indeed short, not allowing for time to form an elliptical resulting from the merger of two or more group members while the system is still recognized as a group. Therefore, when dealing with these compact groups, there is no time to observe the progenitors in different stages of the merging, since the timescale to form the resultant galaxy is very quick. In the case of NGC~1487, we see a pair of objects in interaction, with fully formed tails, while other objects (already perturbed) may still be falling in, and we may hypothesize that they may be at slightly different evolutionary stages, given what we have learned from other similar systems, as described above. However, we believe that only with high resolution HI observations (for better understanding of the gas properties) and numerical simulations, a definitive scenario can be built for this specific system. Indeed, numerical simulations may help constraining the parameters of the interaction for the different progenitors that may be contributing to the present morphological and dynamical status of the NGC~1487 system.

\subsection{On the four bright condensations of NGC~1487}
As mentioned throughout this paper, the four bright condensations, APC1, APC2, APC3 and APC4 proposed by \cite{aguero97} and the tidal tail APC5, have guided the majority of the results of this work (although APC6 and APC7 were also used for the analysis of the kinematics patterns). As a summary, these regions were found to host the youngest ages, highest levels of star formation, the lowest oxygen abundances and the highest electron densities. These regions also revealed to be mostly ionized by star formation and they show the lowest amounts of dust (the higher dust attenuation is actually found in the transitional region between APC1 to APC2). 
In Table \ref{tab:apc} we summarize our results, showing all the derived properties of NGC~1487 for each region and the system as a whole with uncertainties estimated according to section \ref{sec:emissionlines} and adequate propagation of errors.

\begin{table*}
\caption{Physical properties of NGC~1487 and its five main condensations.}
    \hspace*{-0.75cm}
    \ra{1.2}
    \scalebox{0.9}{%
\begin{tabular}{@{}lcccccc@{}} 
\toprule
\textbf{Physical Property} & \textbf{NGC~1487} &  \textbf{APC1} & \textbf{APC2} & \textbf{APC3} & \textbf{APC4} & \textbf{APC5}\\ 
\midrule
Mean SNR (H$\alpha$) & 310 & 482 & 954 & 361 & 556 & 237 \\
f(H$\alpha$) ($10^{-13}$ ergs/s/cm$^2$/\AA) & $16.0 \pm 2.4$ & $3.8 \pm 0.6$ & $7.0 \pm 1.2$ & $2.1 \pm 0.4$ & $1.5 \pm 0.2$ & $0.28 \pm 0.07$   \\
Mean A$_V\text{, gas}$ (mag) & $0.006 \pm 0.001$ & $0.07 \pm 0.02$ & $0.16\pm 0.04$ & $0.09 \pm 0.02$ & $0.05 \pm 0.01$ & $0.08 \pm 0.02$  \\
L(H$\alpha$) (10$^6$ L$_{\odot}$) & $12 \pm 2$ & $2.6 \pm 0.7$ & $5.1 \pm 0.9$ & $1.4 \pm 0.4$ & $0.8 \pm 0.2$ & $0.2 \pm 0.1$ \\
ionization Source  & SF & SF & SF & SF & SF & SF\\
Mean $n_e$ (cm$^{-3}$) & $21 \pm 4$ & $24 \pm 5$ & $17 \pm 3$ & $22 \pm 4$ & $15 \pm 3$ & $12 \pm 2$  \\
SFR (M$_{\odot}$/yr) & $0.37 \pm 0.05$ & $ 0.08 \pm 0.01$ & $0.15 \pm 0.03$ & $0.04 \pm 0.01$ & $0.03 \pm 0.01$ & $0.010 \pm 0.003$ \\

Mean 12 + log(O/H) & $8.33 \pm 0.16$ & $8.33 \pm 0.16$ & $8.29 \pm 0.13$ & $8.30 \pm 0.14$ & $8.27 \pm 0.12$ & $8.27 \pm 0.13$ \\
M$_{\star}$ ($10^9$ M$_{\odot}$) & $4.61 \pm 1.24$ & $1.45 \pm 0.82$ & $0.54 \pm 0.09$ & $0.55 \pm 0.07$ & $0.09 \pm 0.02$ & $0.02 \pm 0.01$ \\
Mean age (light-weighted) (Myr) & $833 \pm 78$ & $989 \pm 82$ & $384 \pm 55$ & $640 \pm 58$ & $287 \pm 46$ & $709 \pm 57$ \\
Mean age (mass-weighted) (Myr) & $1545 \pm 112$ & $1869 \pm 163 $ & $1181 \pm 95$ & $1220 \pm 104$ & $825 \pm 97$ & $1280 \pm 103$ \\
Mean $Z_{\star}$ (light-weighted) & $0.013 \pm 0.007 $ & $0.014 \pm 0.004$ & $0.018 \pm 0.006$ & $0.012 \pm 0.005$ & $0.015 \pm 0.006$ & $0.015 \pm 0.004$ \\
Mean $Z_{\star}$ (mass-weighted) & $0.014 \pm 0.008$ & $0.018 \pm 0.005$ & $0.021 \pm 0.007$ & $0.011 \pm 0.004$ & $0.015 \pm 0.006$ & $0.014 \pm 0.004$ \\
\bottomrule
\end{tabular}}
\label{tab:apc}
\end{table*}

\subsection{Gas-phase metallicity and metal mixing}
In NGC~1487 we detect an overall flat oxygen abundance gradient with large depressions of low-metallicity gas in the central regions of all four condensations, which characterizes an inversion in the oxygen abundance distribution. We observe a systematic shift in the systemic velocity (Fig. \ref{fig:velocity}) from the region of APC1, APC2 and APC4 ($\sim$850 km/s) to the region of APC3 ($\sim$ 750 km/s), indicating that these may be two of the galaxies that are merging (although we can only observe one galactic body nowadays). This transition supports the idea of a gas flow between the galaxies and this gas flow may be the main responsible in producing this inversion in the gas-metallicity gradient. 

Further work using large field-of-view kinematic data with a higher spectral resolution would be a promising path to look for such flows and explain this observed inversion in the metallicity gradient. 

\subsection{Star formation rates} 
The first impression when inspecting all of the previous physical properties derived for this galaxy is the large discrepancy between the emission in the optical, given by the V-band isophotes overplotted to the images, and the nebular emission. There seems to exist an offset between the V-band emission in the Northern region of the galaxy, where the gas emission is nearly zero. In the same context, a strong star forming region is found in the bottom region of APC1.

We conclude that the merging event was responsible for suppressing, or at least not strongly enabling, star formation in the regions that dominate the optical (mainly represented by APC1), but, at the same time, it clearly triggered star formation in the central region of APC2 and APC4.

\subsection{ionization Source} By studying the BPT diagrams, we saw that the majority of the spaxels fall in the star-forming region, being solely ionized by photoionising stars. 

Although studies support that merger events can induce AGN activity, this is not seen in this analysis. However, as \cite{Lee} suggested, this pre-merger is still young (approximately $500$ Myrs), and we can expect that AGN triggering may still happen, probably in the final stages of the merging process.

We identified, in the BPT-SII diagram, signs of LINER-like activity in some spaxels of the system. These spaxels are mostly in the outer-regions of the galaxies and, as shown by \cite{porto3d}, may be reflecting the emission of post AGB stars. These spaxels, however, fall out of the regions we are interested in this work and so, were not further studied.

\subsection{Mass, ages and stellar metallicities} 
The study of the age distribution of the system suggests that the centre of the four condensations host stellar populations that are younger than those present in the outskirts of the galaxies, which is consistent with the higher rates of star formation surface density found in these regions. These younger populations are mostly observed in the light-weighted age map, displaying ages of around $\sim$100 - 300 Myr.

\cite{Mengel}, studying star clusters in the interior of the system, showed that the inner region of APC2 hosts stellar populations with ages of $\sim$10 Myrs. On another hand, \cite{Lee}, performing WFPC2 B- and I-band photometry on 560 clusters of NGC~1487 and its tidal tails, found a roughly bimodal age distribution, peaked at $\sim$15 Myr and $\sim$500 Myr.

In this work, with our pixel-by-pixel estimates of light- and mass-weighted ages, we are able to find, firstly in the light-weighted map, populations younger than 100 Myrs in the region of APC2 and APC4, as well as in the upper region of APC5. In the central regions of APC1 and APC3 the ages seem to converge to the older populations found by \cite{Lee} of $\sim$500 Myrs. In the mass-weighted age map, we see the location of the oldest populations in NGC~1487, mainly found in the region of APC1, reaching $\sim$10 Gyr. 

These findings reinforce the scenario that the system is in a pre-merger stage, as described by \cite{Surace}. In this case, the progenitor galaxies must have had a first contact around $\sim$500 Myrs ago, creating some of its tidal tails and igniting bursts of star formation in the system, generating the older population of star clusters studied by \cite{Lee} and \cite{Mullan}. More recently, the system evolved to a pre-merger stage, being now gravitationally bound and forming the youngest star cluster populations \citep{Mengel}.

\begin{figure}
    \centering
    \includegraphics[width=\columnwidth]{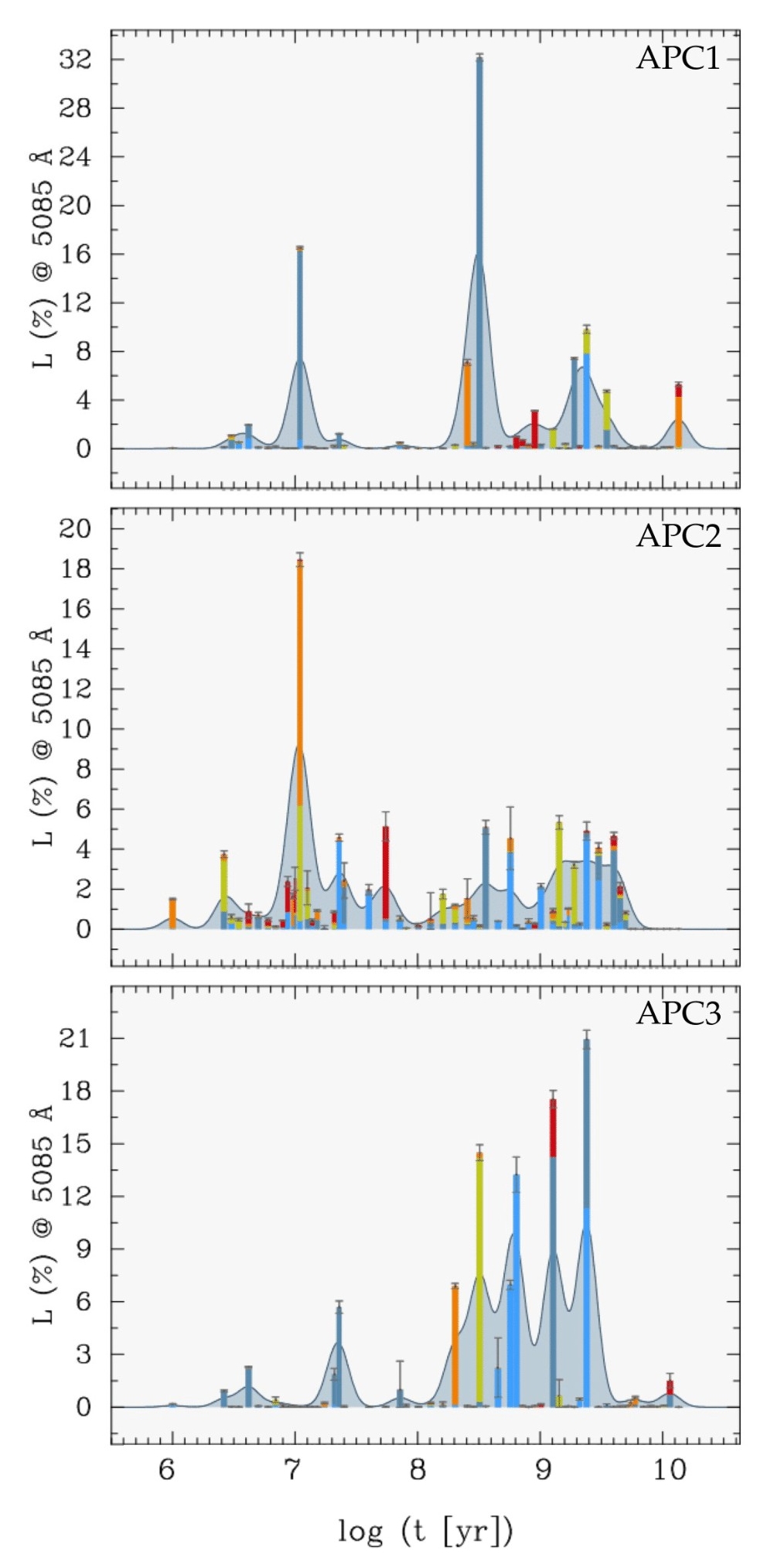}
    \caption{Luminosity fraction at the normalization wavelength (5085 \AA) as a function of the age for the integrated regions of APC1, APC2 and APC3, respectively. The vertical bars represent the $1\sigma$ uncertainties and the colors light-blue, blue, light-green, orange and red stand for metallicities of 4.e-4, 0.004, 0.008, 0.02 and 0.05, respectively. The light-blue shaded area shows an Akima-smoothed \citep{Akima} version of the SSP contributions, giving as a result an illustration of the star formation history of the system. }
    \label{sfh}
\end{figure}

To double-check the stellar populations in the system, we have integrated the flux in the regions APC1, APC2 and APC3 and fitted the resulting spectra with FADO, using an extended SSP library (295 templates; Chabrier IMF \citep{Chabrier} and Padova 1994 stellar tracks \citep{Padova,Padova2}). In Fig. \ref{sfh}, we show the inferred luminosity fraction at the normalization wavelength (5085 \AA), which is a schematic illustration of the star formation history of the system. The peaks seen in the recent past ($\sim 1$ Gyr) imply a relatively strong star formation episode in APC1 at around $\sim$300 Myr ago, one  in APC2 at $\sim$10 Myr ago, and a few star formation episodes since $\sim$1 Gyr in APC3, which is a strong evidence of the bimodal age distribution of NGC~1487, with peaks in $\sim 10$ Myrs and $\sim 300$ Myrs.

\subsection{Kinematics}
From the combination of the velocity field and of the velocity dispersion maps of NGC~1487, we hypothesize that the undergoing merger of the gaseous structure of NGC~1487 will rebuild an unique rotation body if stellar relaxation processes do not destroy it, and this gas rotation will, later on, follow the potential well created by the stellar body of the system.

We compare our results on NGC~1487 with those of the MAD survey \citep{denBrok}. Studying the kinematics of a sample of 41 out of the 45 galaxies within the MAD survey, \cite{denBrok} found that most of the galaxies are axially symmetric and show quite regular velocity fields. Two exceptions are NGC 3256 and NGC 337, which are two merging systems showing small non-circular patterns in the gas and chaotic velocity dispersion distributions. In both these merging systems, however, the gas and stars are mostly coalescent, in great contrast with NGC~1487. Therefore, we did not find them to be good systems to compare with the presently studied object.

The findings of \cite{aguero97}, using  B, V, R, and I photometric and spectroscopic observations of NGC~1487, indicate blue stellar populations in the centre of the system, with an increasing contribution of a reddish component towards the ourskirts. They also show that the regions of APC1, APC2 and APC3 are the ones with highest SFRs, which agrees well with our work. Their study of the kinematics, however, suggests that radial velocities do not show a single disc with circular velocities, while our work highlights the presence of a rotating pattern in the region of APC1, which we attribute to a possible process of rebuilding of a disc.

\subsection{Proposed scenario for NGC 1487} The velocity fields, mass estimates and several physical properties derived in this work led us to the conclusion that NGC 1487 is the result of an ongoing merger event involving smallish dwarf galaxies within a group, in a pre-merger phase. In different aspects, NGC 1487 resembles the Small Magellanic Cloud (SMC). The cold gas component of the SMC shows significant rotation ($\sim$ 60 km/s, \citealt{Stanimirovic}), which is also observed in the warm disc component, while no or little rotation is observed for the old stellar populations \citep{Harris06, Dopita, Suntzeff,Hatzidimitriou}; this implies that the older stellar component is a spheroid that is gravitationally supported by its velocity dispersion. N-body simulations suggest that the kinematical difference between the stellar and gaseous components are due to a major merger event which occurred in the early stage of the galaxy formation \citep{Bekki,Kazantzidis}.  This means that dissipative gas rich dwarf-dwarf merging give rise to a new dwarf, which consists of a spheroidal stellar component and a rotating disc that could have been formed via gas accretion after the formation of the older spheroidal component.  The story is probably different for NGC 1487, as we suggest that this merger event has resulted in the ongoing formation of a dwarf irregular galaxy, given that, despite the very uncommon shape of the stellar velocity field for such a low mass system, we measured significant rotation of the system as a whole, the underlying galaxy which is accreting material from its northern part, could be a dwarf disc galaxy rather than a spheroidal component. Indeed, the gas and stellar components are not in counter-rotation each other, both spin axes appear almost aligned, the merging phase is too advanced and it will probably not be possible to reduce the central angular momentum of the galaxy by moving away some material towards the outskirts of the system by relaxation mechanisms or tidal tail streaming motions.  We suggest that the system will further evolve, into a more passive disc dwarf galaxy. The relic will have a mass similar to that of the SMC, i.e. $5.1 \times 10^9$ M$_{\odot}$ \citep{Harris}. The process of relaxation takes a few orbital times for collisionless stellar systems \citep{Bell}, of the order of $\sim$0.5-2 Gyr, but, due to the fact that energy can be radiated and exchanged through collisions, one may guess that in less than an orbital period the gaseous disc will follow the stellar rotation.
As for the currently observed physical properties of the system, we understand that this upcoming dwarf galaxy will most likely show inverted age and (gas-phase) metallicity gradients, i.e. a central region with younger, less metal-enriched populations than its outskirts, differently from the SMC, for example. 

In order to further understand the current and future morphology of the system, a more detailed study of the gas flows and motions within the main body and two tidal tails of NGC 1487, requires 2D kinematic data over an extended area of the system and with a higher spectral resolution. This is currently being studied using Fabry-Perot data by Torres-Flores et al. in prep.

\section*{Acknowledgements}
We thank Jorge S\'anchez Almeida, for his constructive and helpful comments.

MLB and CMdO are grateful for the support received from the Sao Paulo Research Foundation (FAPESP) grants 2017/50277-0, 2018/09165-6, 2019/01639-1, 2019/23388-0 and 2019/11910-4 and also the support from CAPES. CEB and CMdO also acknowledge the support from FAPESP, grant 2016/12331-0.

This work was supported through FCT grants UID/FIS/04434/2019, UIDB/04434/2020, UIDP/04434/2020 and the project ``Identifying the Earliest Supermassive Black Holes with ALMA (IdEaS with ALMA)'' (PTDC/FIS-AST/29245/2017).

Based on observations collected at the European Organisation for Astronomical Research in the Southern Hemisphere under ESO programme(s) 0100.B-0116(A).This research has made use of the NASA/IPAC Extragalactic Database (NED), which is operated by the Jet Propulsion Laboratory, California Institute of Technology, under contract with the National Aeronautics and Space Administration.

\textit{Additional software}: Astropy \citep{2013A&A...558A..33A, 2018AJ....156..123A}, , matplotlib \citep{4160265}, numpy \citep{5725236}, scipy \citep{scipy}.

\section{Data availability}
The data are available via the ESO Science Archive Facility (archive.eso.org) at the Science Portal (Processed Data) tab.

\bibliographystyle{mnras}
\bibliography{paper.bib}

\bsp	
\label{lastpage}
\end{document}